\pdfoutput=1

\documentclass[11pt,twoside,a4paper,cmspaper,final,collab]{cms-tdr}
\def\svnVersion{315674}\def\cmsCernNo{2013-037}\def\cmsCernDate{\today}\def\cmsMessage{Published in Physics Letters B as \href{http://dx.doi.org/10.1016/j.physletb.2015.11.042}{\doi{10.1016/j.physletb.2015.11.042.}}}

\begin{document}\cmsNoteHeader{HIG-13-024}

\hyphenation{had-ron-i-za-tion}
\hyphenation{cal-or-i-me-ter}
\hyphenation{de-vices}
\RCS$Revision: 308138 $
\RCS$HeadURL: svn+ssh://svn.cern.ch/reps/tdr2/papers/HIG-13-024/trunk/HIG-13-024.tex $
\RCS$Id: HIG-13-024.tex 308138 2015-11-02 02:06:56Z alverson $
\newlength\cmsFigWidth
\ifthenelse{\boolean{cms@external}}{\setlength\cmsFigWidth{0.95\columnwidth}}{\setlength\cmsFigWidth{0.8\textwidth}}
\ifthenelse{\boolean{cms@external}}{\providecommand{\cmsLeft}{top}\xspace}{\providecommand{\cmsLeft}{left}\xspace}
\ifthenelse{\boolean{cms@external}}{\providecommand{\cmsRight}{bottom}\xspace}{\providecommand{\cmsRight}{right}\xspace}
\newcommand{\PA}{\ensuremath{\cmsSymbolFace{A}}\xspace}
\providecommand{\FEYNHIGGS}{\textsc{feynhiggs}\xspace}
\cmsNoteHeader{HIG-13-024}
\title{Search for neutral MSSM Higgs bosons decaying to $\PGmp\PGmm$ in pp collisions at $\sqrt{s}=7$ and 8\TeV }

\date{\today}

\abstract{A search for neutral Higgs bosons predicted in the minimal supersymmetric standard model (MSSM) for $\PGmp\PGmm$ decay channels is presented. The analysis uses data collected by the CMS experiment at the LHC in proton-proton collisions at centre-of-mass energies of 7 and 8\TeV, corresponding to integrated luminosities of 5.1 and 19.3\fbinv, respectively. The search is sensitive to Higgs bosons produced either through the gluon fusion process or in association with a $\bbbar$ quark pair. No statistically significant excess is observed in the $\PGmp\PGmm$ mass spectrum. Results are interpreted in the framework of several benchmark scenarios, and the data are used to set an upper limit on the MSSM parameter $\tan\beta$ as a function of the mass of the pseudoscalar \PA boson in the range from 115 to 300\GeV. Model independent upper limits are given for the product of the cross section and branching fraction for gluon fusion and b quark associated production at $\sqrt{s}=8\TeV$. They are the most stringent limits obtained to date in this channel.}

\hypersetup{%
pdfauthor={CMS Collaboration},
pdftitle={Search for neutral MSSM Higgs bosons decaying to mu+mu- in pp collisions at sqrt(s) = 7 and 8 TeV},%
pdfsubject={CMS},%
pdfkeywords={CMS, physics, Higgs, MSSM, dimuons}}

\maketitle
\section{Introduction}
\label{par:Intro}
The predictions of the standard model (SM)~\cite{SM1,SM2,SM3,H1,H2,H3,H4} of fundamental interactions have been
confirmed by a large number of experimental measurements. The observation of a new
boson with a mass of 125\GeV and properties compatible with those of the SM Higgs boson~\cite{Higgs_atlas,newboson,newboson2}, confirms the mechanism of the electroweak symmetry breaking (EWSB). Despite the success of this theory in describing the phenomenology of particle physics at present collider energies,
the mass of the Higgs boson in the SM is not protected against quadratically divergent quantum-loop corrections at high energy. Supersymmetry (SUSY)~\cite{SUSY1,SUSY2} is one example of alternative models that address this problem. In SUSY, such divergences are cancelled by introducing a symmetry between fundamental bosons and fermions.

The minimal supersymmetric extension of the standard model (MSSM)~\cite{mssm_1,mssm_2} predicts the existence of two Higgs doublet fields. One doublet couples to up-type and one to down-type fermions. After EWSB, five physical Higgs bosons remain: a CP-odd neutral scalar \PA, two charged scalars \Hpm, and two CP-even neutral scalar particles \Ph and \PH. The neutral bosons \Ph, \PA, and \PH, will be generically referred to as $\phi$ collectively in this paper, unless differently specified.

At lowest order in perturbation theory, the Higgs sector in the MSSM can be described in terms of two free parameters: $m_{\PA}$, the mass of the neutral pseudoscalar A, and $\tan\beta$, the ratio of the vacuum expectation values of the two Higgs doublets. The masses of the other four Higgs bosons can be expressed in terms of these two parameters and other measured quantities, such as the masses $m_{\PW} $ and $m_{\Z}$ of the $\PW$ and \Z bosons, respectively.
In particular, the masses of the neutral MSSM scalar Higgs bosons \PH and \Ph are given~\cite{mssm_1} by
\ifthenelse{\boolean{cms@external}}{
\begin{multline}
m_{\PH,\Ph} = \Biggl[\frac{1}{2}\Bigl\{m^2_{\PA}+m^2_{\Z} \pm \bigl[(m^2_{\PA}+m^2_{\Z})^2 \\
- 4m^2_{\PA}m^2_{\Z}\cos^2 2\beta\bigr]^{1/2}\Bigr\}\Biggr]^{1/2}.
\label{eq:Higgsmass}
\end{multline}
}{
\begin{equation}
m_{\PH,\Ph} = \left[\frac{1}{2}\left\{m^2_{\PA}+m^2_{\Z} \pm \bigl[(m^2_{\PA}+m^2_{\Z})^2 - 4m^2_{\PA}m^2_{\Z}\cos^2 2\beta\bigr]^{1/2}\right\}\right]^{1/2}.
\label{eq:Higgsmass}
\end{equation}
}
The \PA and \PH bosons are degenerate in mass above 140\GeV and for small $\cos\beta$ (large $\tan\beta$) values.
This expression also provides an upper bound on the mass of the light scalar Higgs boson, corresponding to $m_{\Ph} \leq m_{\Z} \abs{\cos 2\beta}$. The value can become as large as $m_{\Ph} \approx 135\GeV$ once radiative corrections are taken into account~\cite{Degrassi}.

\begin{figure}[!ht]
\centering
\includegraphics[width=0.48\textwidth]{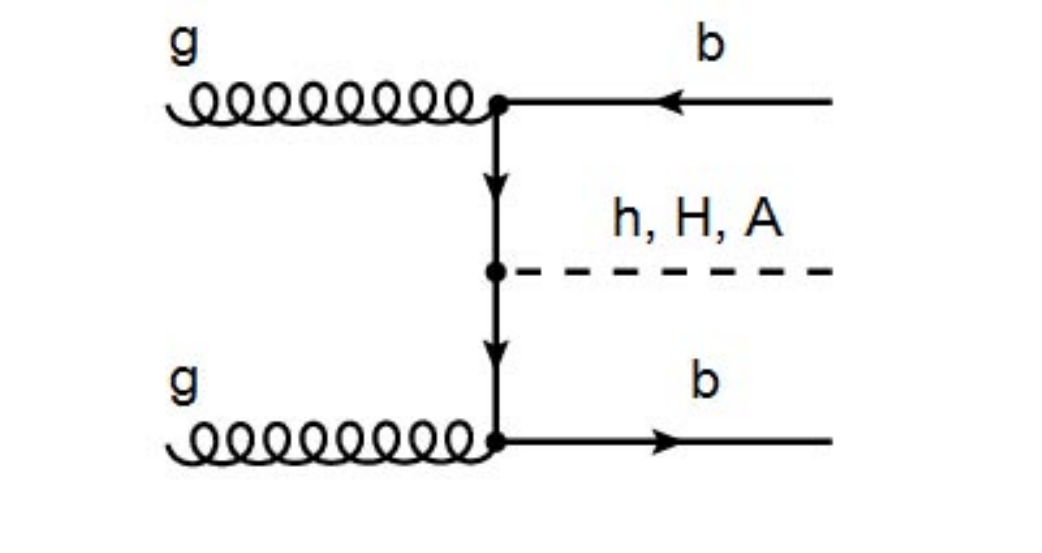}\label{fig:bbprod}
\includegraphics[width=0.48\textwidth]{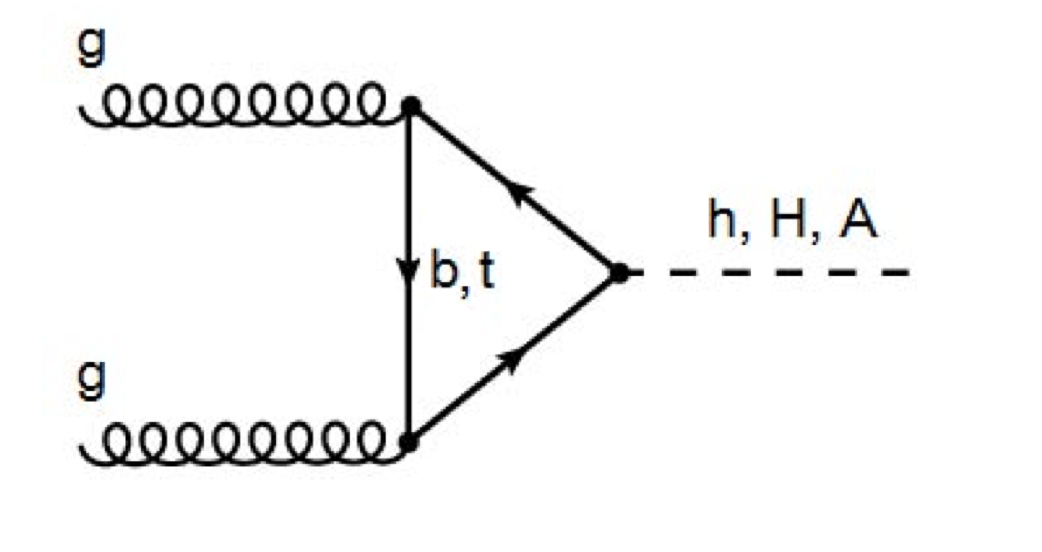}\label{fig:ggprod}
\caption{Leading-order diagrams for the main production processes of MSSM Higgs bosons at the LHC (\cmsLeft) in association with $\bbbar$ production and (\cmsRight) through gluon fusion.}
\label{fig:MSSM_higgs}

\end{figure}

The main production mechanisms for the three neutral $\phi$ bosons at the LHC are
the associated production with $\bbbar$ quarks (AP), given at the leading order by the
Feynman diagram shown in
Fig.~\ref{fig:MSSM_higgs}\,(\cmsLeft), and the gluon fusion (GF) process, shown
in Fig.~\ref{fig:MSSM_higgs}\,(\cmsRight)~\cite{cs_higgs1,cs_higgs2,cs_higgs3}. The GF process with virtual t or b quarks in the loop is dominant at small and moderate values of $\tan\beta$. At large  $\tan\beta$ the coupling of $\phi$ to down-type quarks is enhanced relative to the SM~\cite{MSSM_mhmax_top} and the AP process becomes dominant. Similarly, the coupling of the $\phi$ boson to charged leptons is also enhanced at large $\tan\beta$.

This paper reports on a search for the MSSM neutral Higgs bosons produced either by the AP or GF mechanisms, where the Higgs bosons decay via $\phi \to \PGmp\PGmm$. The analysis is sensitive to all the three bosons, \Ph, \PH, and \PA in the mass range between 115 and 300\GeV.
The search is performed by the CMS collaboration using
data recorded in pp collisions at the LHC, corresponding to an integrated luminosity of 5.1\fbinv at $\sqrt{s}=7\TeV$ and 19.3\fbinv at $\sqrt{s}=8\TeV$. The common experimental signature of the two processes is a pair of
oppositely charged muons with high transverse momentum (\pt) and a small imbalance of \pt in the event.
The AP process is characterized by the presence of additional jets originating from b quarks (b jets), whereas the events with only jets from light quarks or gluons are sensitive to the GF production mechanism. The presence of a signal would be characterized by an excess of events over the background in the dimuon invariant mass corresponding to the $\phi$ mass value.

Although the product of the cross section and the branching fraction for the $\PGmp\PGmm$ channel is a factor $10^3$ smaller than
for the corresponding $\tau^+ \tau^-$ final state, the muon pair can be fully reconstructed,
and the invariant mass precisely measured by exploiting the excellent muon momentum resolution of the CMS detector. Searches for the MSSM Higgs bosons have been performed at LHC by the LHCb experiment in the $\tau^+ \tau^-$ final state at large pseudorapidity values~\cite{MSSM_LHCb}, the ATLAS experiment in the $\PGmp\PGmm$ and $\tau^+ \tau^-$ channels~\cite{ATLAS-MSSM2,ATLAS-MSSM3}, and by the CMS experiment in the $\tau^+ \tau^-$~\cite{MSSM_Htautau} and $\bbbar$~\cite{MSSM_Hbb, MSSM_hbb2} final states. Limits on the existence of MSSM Higgs bosons were also determined at Tevatron~\cite{CDF_1,CDF_2,D0_1,D0_2} and at LEP~\cite{LEP-MSSM}.

Traditionally, searches for MSSM Higgs bosons are presented in the context of benchmark scenarios that describe the mass relation
among the three neutral MSSM Higgs bosons, their widths, and cross sections. Each scenario assigns well defined values to the relevant parameters of the MSSM, except $m_{\PA}$ and $\tan\beta$, which are left free to vary. The $m_{\Ph}^\text{max}$ benchmark scenario~\cite{MSSMscenario,MSSM_mhmax_top} provides $m_{\Ph}$ values as large as 135\GeV, and the weakest bounds on $\tan\beta$ for fixed values of the top quark mass. For this reason, it has been used in most of
the previously quoted analyses to present the results from MSSM Higgs boson searches. However, within the MSSM the newly discovered state with a mass of 125\GeV can be interpreted as the light CP-even Higgs boson, \Ph~\cite{H125-MSSM1}. In this case,
a large part of the $m_{\PA}$--$\tan\beta$ parameter space is excluded within the $m_{\Ph}^\text{max}$ scenario,
and new benchmarks were therefore proposed in which the MSSM parameters are adjusted to have $m_{\Ph}$ in the interval 122 to 128\GeV, but with a wider range of $\tan\beta$ and $m_{\PA}$ values~\cite{MSSMscenario,MSSM_mhmax_top,H125-MSSM1}.
To do this, the $m_{\Ph}^\text{max}$ scenario was reformulated in two versions, $m_{\Ph}^\mathrm{mod +}$ and $m_{\Ph}^\mathrm{mod -}$, corresponding to different values of the top squark mixing parameter.
Other recently proposed scenarios~\cite{MSSMscenario} are the light top squark (light stop) model, which results in a modified GF rate, and the light tau slepton (light stau) model,
which yields a modified ${\Ph} \to \gamma \gamma$ branching fraction. Such models are expected mainly to affect
the Higgs boson production cross section and not the kinematic properties of the events. A list of the parameters of the various scenarios can be found in Ref.~\cite{MSSM_Htautau}.
The results presented in this paper are obtained in the framework of the MSSM $m_{\Ph}^\mathrm{mod +}$ scenario. Comparisons are also made with other benchmarks.

\section{The CMS detector and event reconstruction}
\label{par:Det}

The central feature of the CMS apparatus is a superconducting solenoid of 6\unit{m} internal diameter, providing a magnetic field of 3.8\unit{T}. Within the solenoid volume are a
silicon pixel and strip tracker, a lead tungstate crystal electromagnetic calorimeter (ECAL), and
a brass and scintillator hadron calorimeter (HCAL), each comprised of a barrel and two endcap
sections. Muons are measured in gas-ionization detectors embedded in the steel flux-return
yoke outside the solenoid. Forward calorimetry extends the coverage provided
by the barrel and endcap detectors up to pseudorapidity $\abs{\eta}<5$.
A detailed description of the CMS detector, together with a definition of the coordinate system and kinematic variables, can be found in Ref.~\cite{CMSdetector}.
The CMS offline event reconstruction creates a global event description using the particle flow (PF) technique~\cite{particleflow}. The PF event reconstruction attempts to reconstruct and identify each particle with an optimized combination of all
subdetector information. The missing \pt vector is defined as the projection on the plane perpendicular to the beams of the negative vector sum of the momenta of all reconstructed particles in an event. Its magnitude is referred to as \ETm.

An average of 9 and 21 pp collisions take place in any LHC bunch crossing, respectively at 7 and 8\TeV, because of the large luminosity of the machine and the size of the total inelastic cross section. These overlapping events (pileup) are characterized by small-\pt tracks, compared to the particles produced in a $\phi \to \PGmp\PGmm$ event, and their presence can degrade the detector capability to reconstruct the objects relevant for this analysis. The primary vertex is chosen from all reconstructed interaction vertices as the one with the largest sum in the squares of the \pt of the associated tracks.
The charged tracks originating from another vertex are then removed.

Offline jet reconstruction is performed using the anti-\kt
clustering algorithm~\cite{Cacciari:2008gp, Cacciari:2011ma} with a distance parameter of 0.5.
The jet momentum is defined by the vectorial sum of all the PF particles momenta in the jet, and found in simulation
to be within 5\% to 10\% of the true hadron-level momentum, with some \pt and $\eta$ dependence. Extra energy coming from pileup interactions affects the momentum measurement. Corrections to the measured jet energy are therefore applied. They are derived from event simulation, and confirmed with in-situ measurements using energy balance in dijet and \Z/photon+jet events~\cite{jes}.

Muons are measured in the pseudorapidity range $\abs{\eta}< 2.4$, using detection planes based on three technologies: drift tubes, cathode strip chambers, and resistive-plate chambers. Matching muons to tracks measured in the silicon tracker provides relative \pt resolutions for muons with $20<\pt<100\GeV$ of 1.3--2.0\% in the barrel and better than 6\% in the endcaps. The $\pt$ resolution in the barrel is better than 10\% for muons with \pt up to 1\TeV~\cite{MuonPOG_efficSys}.

\section{Simulated samples}
\label{sec:Data}

Simulated samples are used to model the signal and to determine the efficiency of the signal selection. Background samples are also simulated to optimize the selection criteria. The normalization and distribution of the background events are measured from data.

The signal samples are generated using the Monte Carlo (MC) event generator
\PYTHIA 6.424~\cite{pythia} for a wide range of $m_{\PA}$ and $\tan\beta$ values, as listed in Table~\ref{tab:MC_signal},
for the AP and the GF production mechanisms.
The $\phi$ production cross sections and their corresponding uncertainties are provided by the
LHC Higgs Cross Section Working Group~\cite{cs_higgs1,cs_higgs2,cs_higgs3}.
The cross sections for the
GF process in the $m_{\Ph}^\text{max}$ scenario are obtained using the HIGLU program~\cite{higlu1,higlu2}, based on next-to-leading order (NLO) quantum chromodynamics (QCD) calculations. The \textsc{sushi} program~\cite{sushi} is used for the other benchmarks. For the AP process, the four-flavor NLO QCD calculation~\cite{bbphi__1,bbphi__2}
and the five-flavor next-to-next-to-leading order (NNLO) QCD calculation are implemented in \textsc{bbh@nnlo}~\cite{bbhnnlo} and combined using the Santander matching scheme~\cite{santander}.
The Higgs Yukawa couplings computed with the
\FEYNHIGGS program~\cite{Feynhiggs} are used in the calculations. The decay branching fractions to muons in the different
benchmark scenarios are obtained with \FEYNHIGGS and \HDECAY~\cite{hdecay}. Further details on signal generation can be found in Refs.~\cite{cs_higgs1,cs_higgs2,cs_higgs3}.

The values of $m_{\Ph}$ predicted by \FEYNHIGGS differ typically by a few \GeV{} from those computed with \PYTHIA. The invariant mass spectrum of the \Ph boson is therefore shifted to match the \FEYNHIGGS prediction.
The small difference between \PYTHIA and \FEYNHIGGS in assessing the width of the h boson is
of the order of 100\MeV, and therefore neglected, since the experimental mass resolution is at least one order of magnitude larger. The \PYTHIA parameters used to simulate the signal are those for the $m_{\Ph}^\text{max}$ scenario. Since for a given set of $m_{\PA}$ and $\tan\beta$ values,
the kinematic properties of the final state are the same for all the scenarios, the simulated samples based on the $m_{\Ph}^\text{max}$ benchmark are also used to check the validity of the other models. Further details on this procedure and the related systematic uncertainties are discussed in Section~\ref{sec:sys}.

\begin{table}[!ht]
\topcaption{The $m_{\PA}$ and $\tan\beta$ values used to generate signal samples.}
\centering
\begin{tabular}{  c  c  c  c }
\hline
$m_{\PA}$ (\GeVns{})    & $m_{\PA}$ step (\GeVns{}) & $\tan\beta$ & $\tan\beta$ step   \\ \hline
115--200        &  5                         & 5--55             & 5                         \\
200--300        &  25                        & 5--55             & 5                         \\
300--500        &  50                        & 5--55             & 5                         \\
\hline
\end{tabular}

\label{tab:MC_signal}
\end{table}

The main source of background for the $\phi$ production and decay to $\PGmp\PGmm$
is Drell--Yan muon-pair production, $\qqbar \to Z/\gamma^{*} \to \PGmp\PGmm$. Another background is from oppositely charged muon pairs produced in decays of top quarks in $\ttbar$ production. These events are simulated using the \MADGRAPH5.1~\cite{madgraph} generator.  Other background processes such as $\PW^{\pm}\PW^{\mp}$, $\PW^{\pm}\Z$, and $\Z\Z$ are generated with \PYTHIA. The MC samples also include simulated pileup events to reproduce the overlapping pp interactions present in the data.
All generated events are processed through a detailed simulation of the CMS
detector based on \GEANTfour~\cite{geant} and are reconstructed with the same algorithms used for data.

\section{Event selection}
\label{par:EvSel}
The experimental signature of the MSSM Higgs bosons decay considered in this analysis
is a pair of oppositely charged muons with high \pt.
The invariant mass of the pair corresponds to the mass of the $\phi$ boson within the experimental resolution. Moreover, the process is characterized by a small $\ETm$ in the event. If the $\phi$ boson is produced in association with a $\bbbar$ pair, the presence of at least one b quark jet is expected.

\begin{figure}[!htb]
\centering
\includegraphics[width=0.48\textwidth]{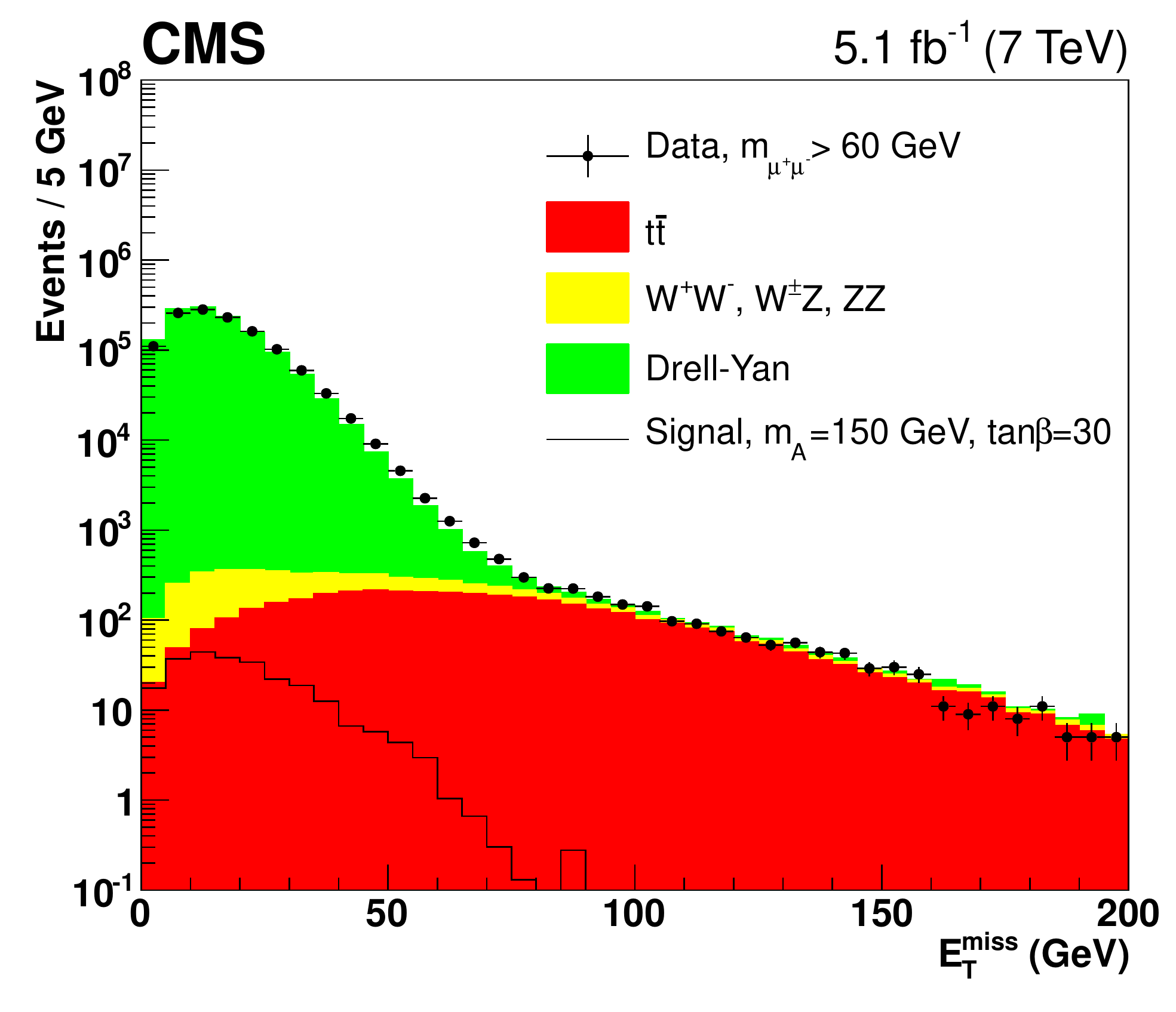}\label{fig:Etmiss2011}
\includegraphics[width=0.48\textwidth]{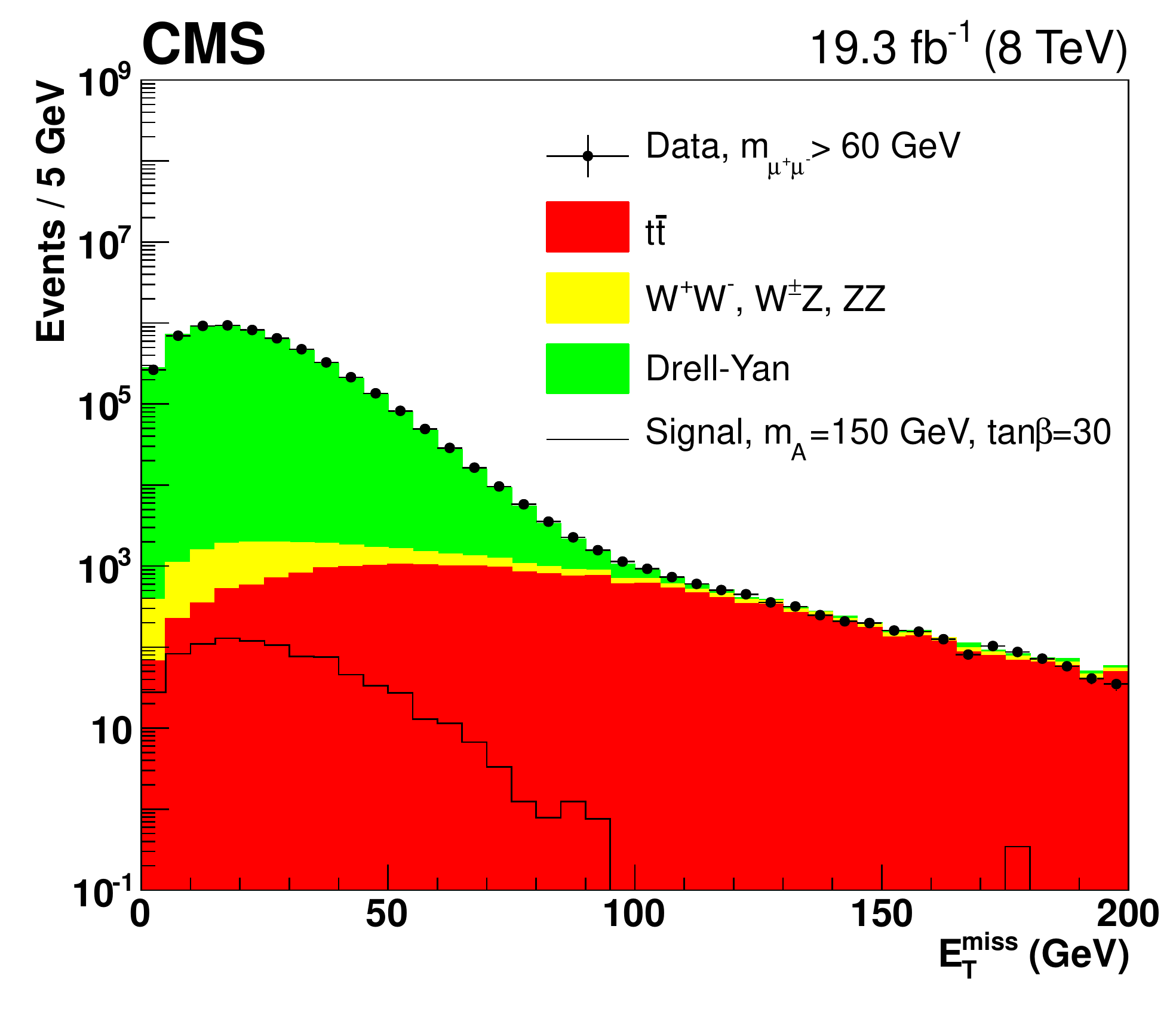}\label{fig:Etmiss2012}
\caption{The $\ETm$
distribution for events with a reconstructed dimuon invariant mass $m_{\PGmp\PGmm}>60\GeV$ in data and in simulated events at $\sqrt{s}=7$ (\cmsLeft) and  $\sqrt{s}=8\TeV$ (\cmsRight). The expected contribution is also shown for a signal at $m_{\PA}=150\GeV$ and $\tan\beta=30$.}
\label{fig:Etmiss}

\end{figure}

\begin{figure}[!htb]
\centering
\includegraphics[width=0.48\textwidth]{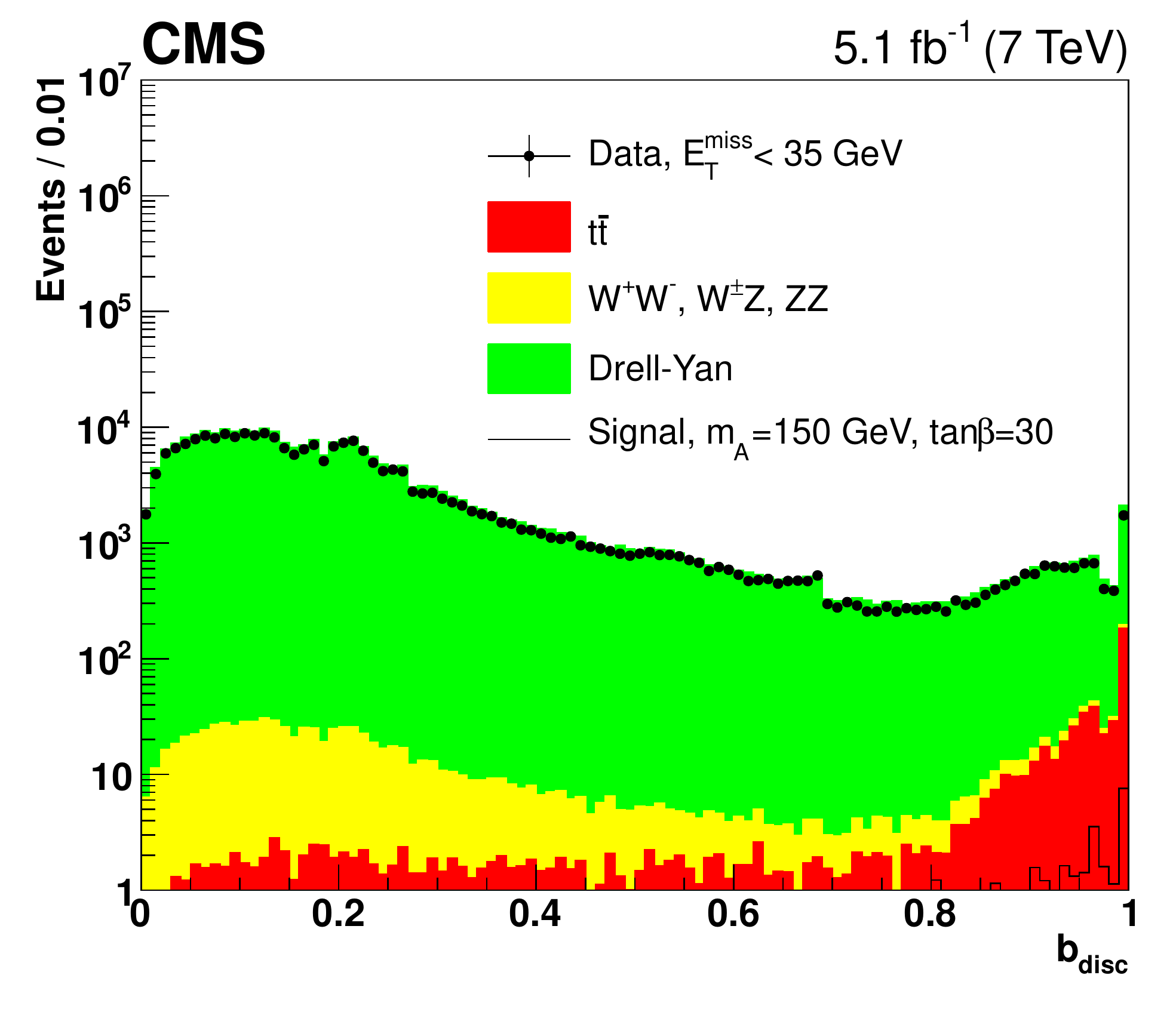}\label{fig:bdisc2011}
\includegraphics[width=0.48\textwidth]{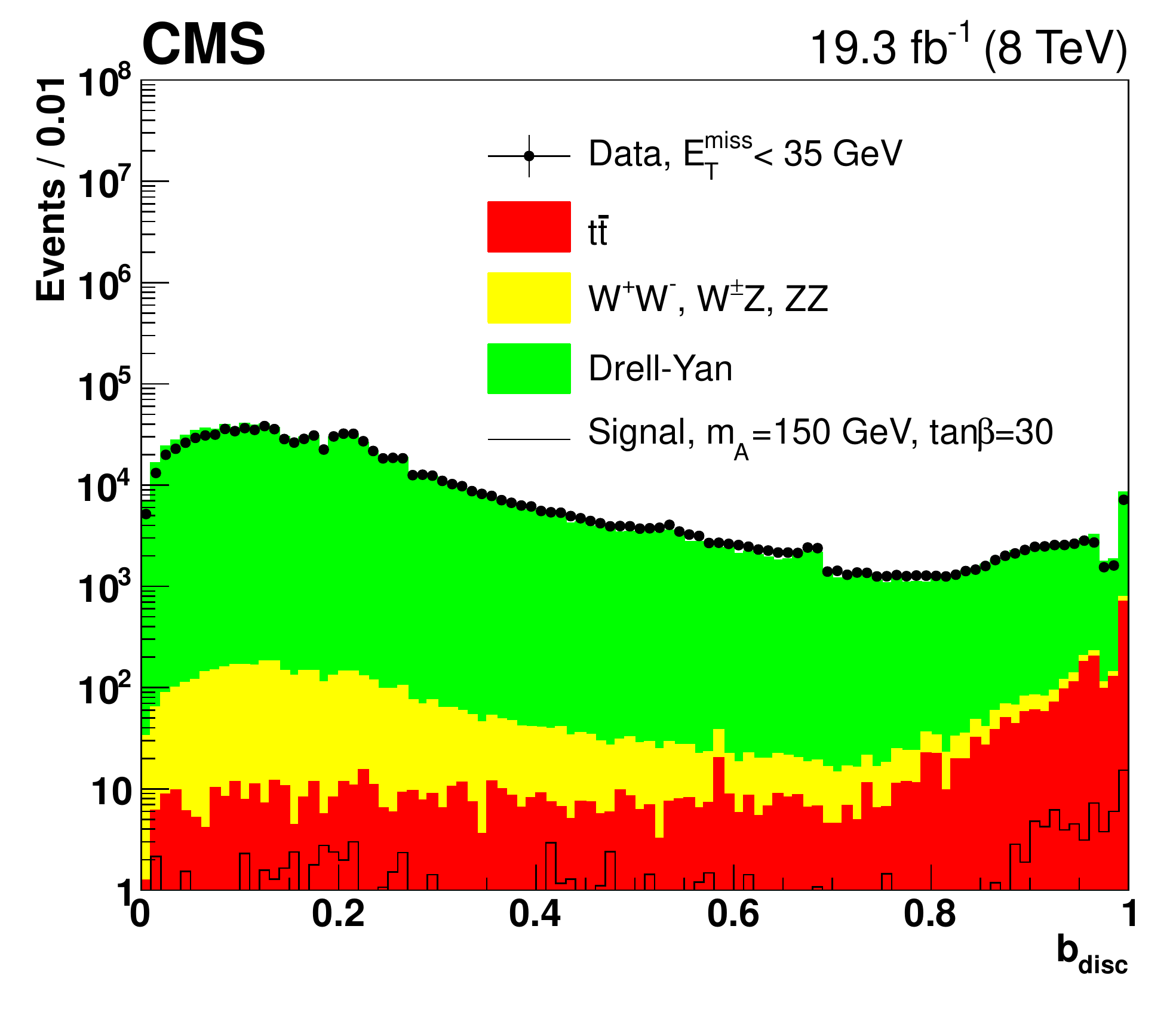}\label{fig:bdisc2012}
\caption{The distribution of the b tagging discriminant, $b_\mathrm{disc}$,  for events that satisfy the selection $\ETm < 35\GeV$ in data collected at $\sqrt{s}=7$ (\cmsLeft) and  $\sqrt{s}=8\TeV$ (\cmsRight). For each event, the largest value of $b_\mathrm{disc}$ is selected.}
\label{fig:bdisc}

\end{figure}
The details of the event selection are listed below, and summarized in  Table~\ref{tab:EventSelection}.
The events are selected using a single-muon trigger, which requires at least one isolated muon with $\pt >24\GeV$ in the pseudorapidity range $\abs{\eta}<2.1$.  The distance of the primary vertex along the $z$ axis from the nominal centre of the detector must be $\abs{z_\mathrm{PV}}<24$\unit{cm}. Muon candidates are reconstructed and identified using both the inner tracker and the muon detector information.
The selected events must have at least two oppositely-charged muon candidates, each with $\pt > 25\GeV$. In events with more than two muon candidates, the two with opposite charges and the highest \pt are retained. The $\eta$ of both muon candidates is chosen to match the trigger acceptance.
Each muon track must have at least one hit in the pixel detector, more than five or eight layers with hits in the tracker, respectively, for the 8 and 7\TeV data
and a directional matching to hits in at least two different muon detector planes. In addition the global fit to the hits of the muon candidate must include at least one hit in the muon detector. The $\chi^2/\mathrm{dof}$ of the global fit of the muon track must be smaller than 10. These requirements ensure a good measurement of the momentum, and significantly reduce the amount of hadronic punch-through background~\cite{MuonPOG_efficSys}. To reject cosmic ray muons, the transverse and longitudinal impact parameters of each muon track must satisfy the requirements $\abs{d_{xy}} < 0.02$\unit{cm} and $\abs{d_{z}} < 0.1$\unit{cm}, respectively.
Both parameters are defined relative to the primary vertex. To ensure that the trigger muon candidate is well-matched to the reconstructed muon track, at least one of the two muon tracks is required to match the direction of the trigger candidate within a cone $\Delta R = 0.2$, where $\Delta R = \sqrt{\smash[b]{(\Delta \eta) ^2 + (\Delta \varphi)^2}}$ is the distance between the muon track and the trigger candidate direction in the $\eta$--$\varphi$ plane, with $\varphi$ being the azimuthal angle measured in radians. Both reconstructed muon candidates must fulfill isolation criteria. A muon isolation variable is constructed using the scalar sum of the \pt of all PF particles, except the muon, reconstructed within a cone \mbox{$\Delta R = 0.4$} around the muon direction. A correction is applied to account for the possible contamination from neutral particles arising from pileup interactions.
A muon is
accepted if the value of the corrected isolation variable is less than 12\% of the muon $\pt$.

\begin{table}[!ht]
\topcaption{Event selection: the criteria listed in the upper part of the table are common to the C1 and C2 categories, that are then mutually exclusive.}
\centering
\begin{tabular}{  c  c  }
\hline
 \multicolumn{2}{c}{Common selection} \\
\hline
Single muon trigger &  $\pt > 24\GeV$ + isolation + $\abs{\eta}<2.1$ \\
Event primary vertex & $\abs{z_\mathrm{PV}}<24$\unit{cm} \\
Muon selection & 2 opposite-charged muons, \\
               & $\pt > 24\GeV$, $\abs{\eta}<2.1$, \\
               &  track quality cuts, \\
               & $\abs{d_{xy}} < 0.02$\unit{cm}, $\abs{d_{z}} < 0.1$\unit{cm}, \\
               & angular matching with trigger, \\
               & isolation \\
 $\ETm$       &  $\ETm < 35\GeV$ \\
\hline
\hline
 \multicolumn{2}{c}{Category C1} \\
\hline
b tag    & 1 or 2 b-tagged jets, \\
         & $\pt^\text{jet}> 20\GeV$, $\abs{\eta^\text{jet}}<2.4$ \\
\hline
\hline
 \multicolumn{2}{c}{Category C2} \\
\hline
No b tag & Events with no b-tagged jets \\
\hline

\hline

\end{tabular}

\label{tab:EventSelection}
\end{table}

A selection based on $\ETm$ provides good separation between signal events and $\ttbar$ background, in the case of leptonic decay of the \PW~boson from top decay.
 The $\ETm$ distributions for events collected at $\sqrt{s}=7$ and 8\TeV are shown
in Fig.~\ref{fig:Etmiss} for events with a reconstructed muon pair with invariant mass
$m_{\PGmp\PGmm}>60\GeV$.
The background contributions from SM processes are superimposed.
For illustration, the expected distribution for signal processes is also shown for $m_{\PA}=150\GeV$ and $\tan\beta=30$.
Studies performed using the simulation show that the $\ETm$ distribution for signal events does not vary significantly for different $m_{\PA}$ and $\tan\beta$ assumptions, and indicate that the selection $\ETm < 35\GeV$
provides highest sensitivity for signal at both centre-of-mass energies.

\begin{figure*}[!ht]
\centering
\includegraphics[width=0.48\textwidth]{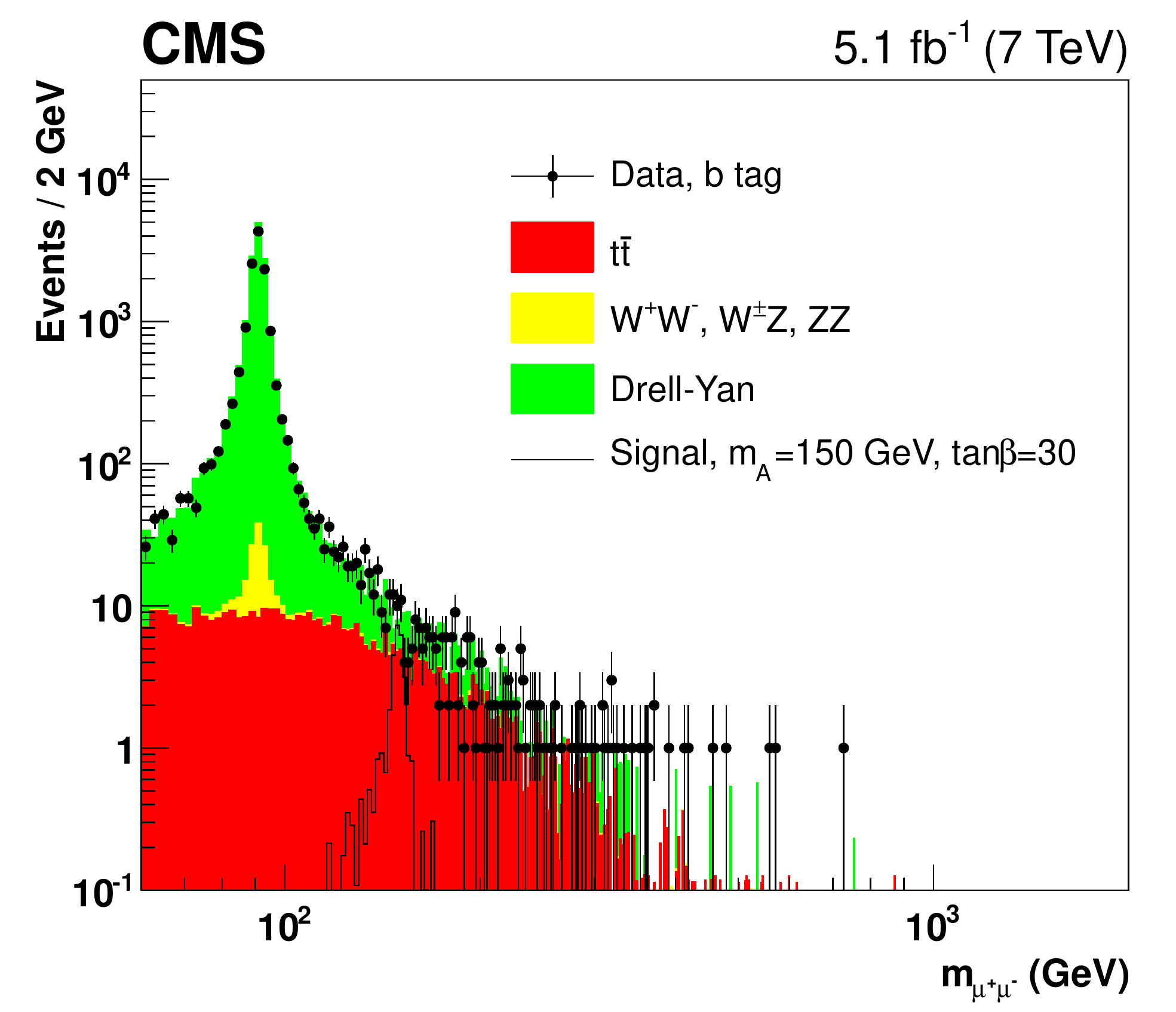}\label{fig:minv_cat1_2011}
\includegraphics[width=0.48\textwidth]{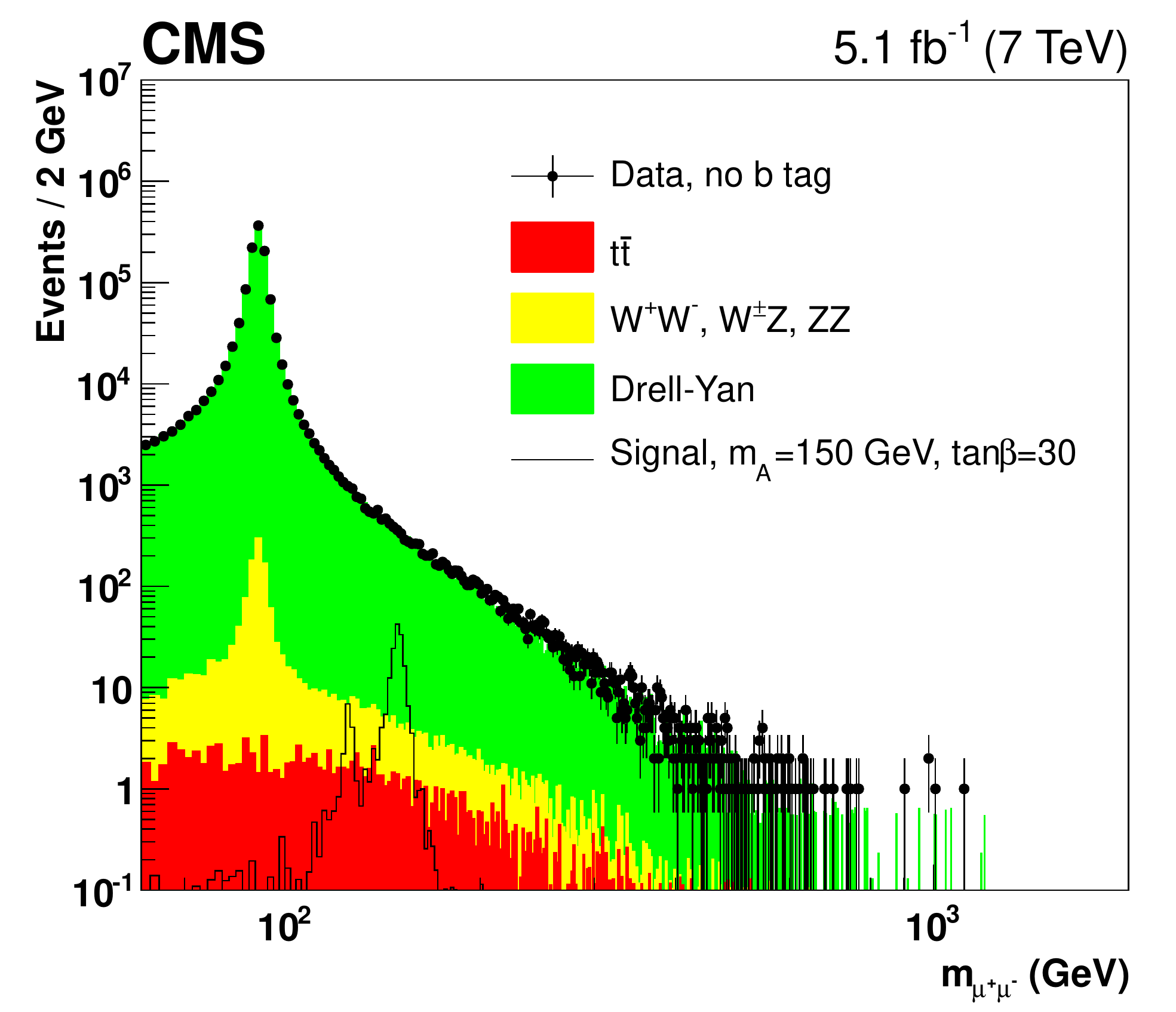}\label{fig:minv_cat2_2011}
\includegraphics[width=0.48\textwidth]{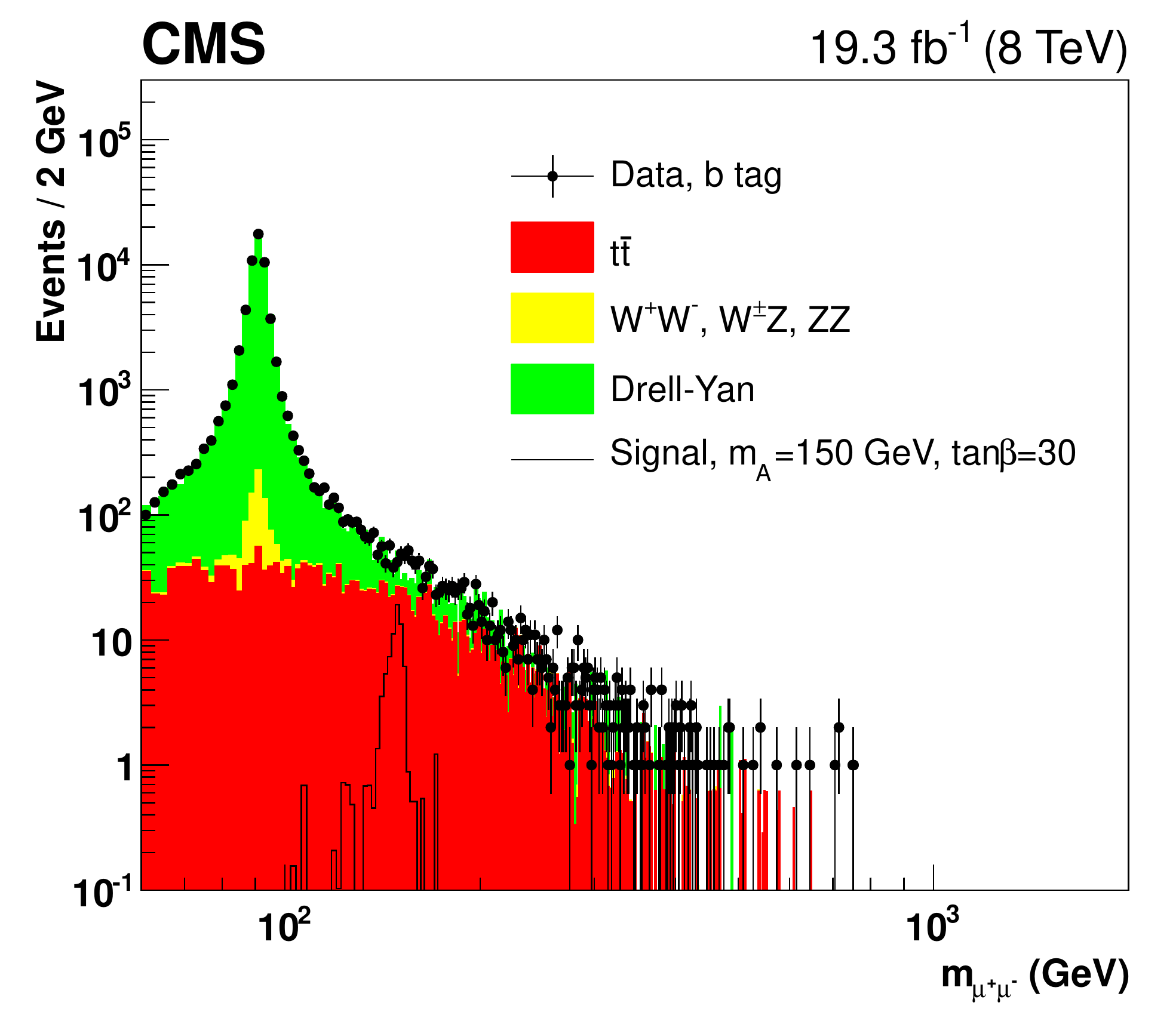}\label{fig:minv_cat1_2012}
\includegraphics[width=0.48\textwidth]{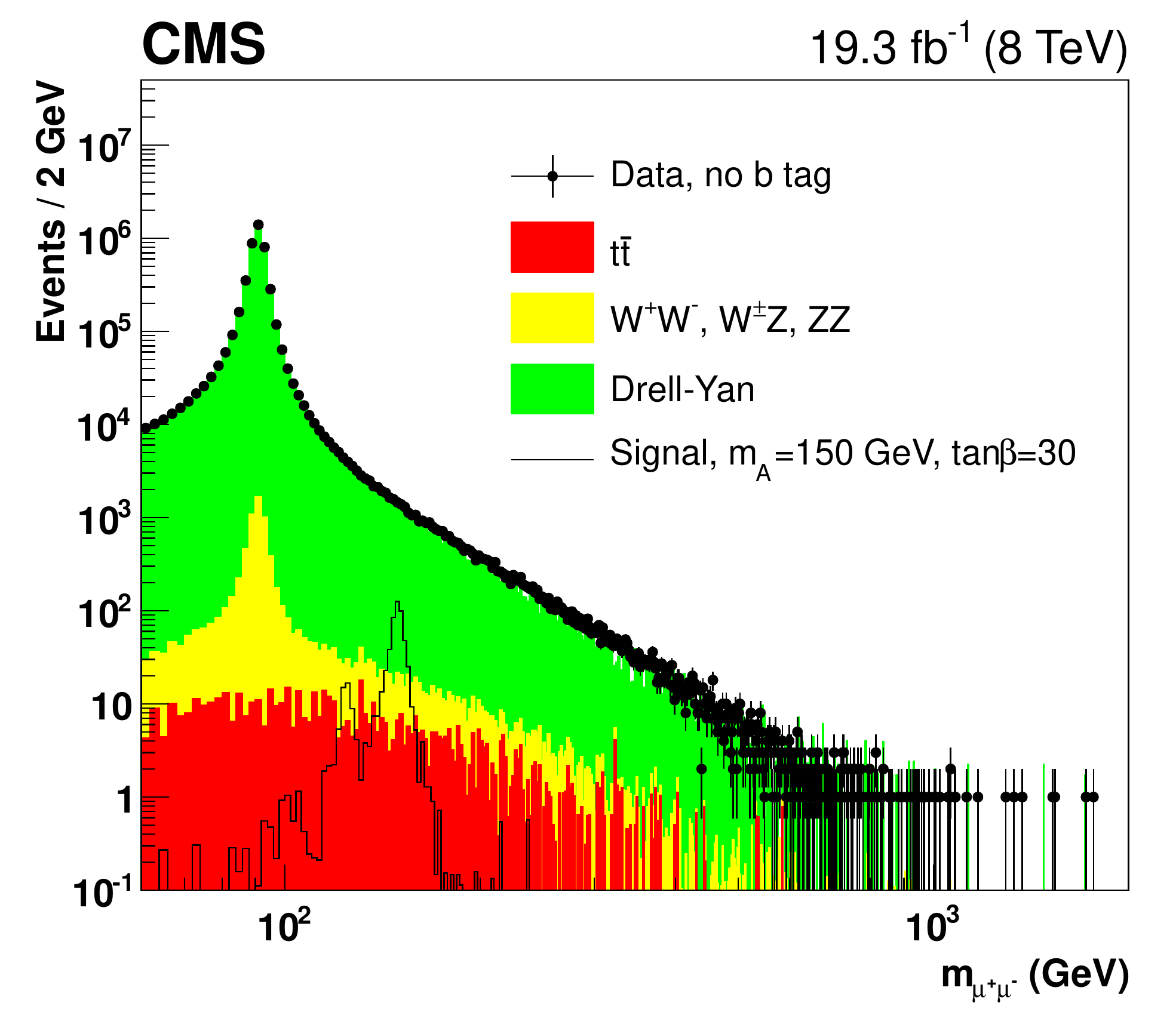}\label{fig:minv_cat2_2012}
\caption{The dimuon invariant mass distribution for events that belong to C1 (upper left) and C2 category (upper right), for data and simulated events at $\sqrt{s}=7\TeV$. The corresponding quantities are shown for $\sqrt{s}=8\TeV$ (lower left and lower right).  The expected contributions to signal assuming the $m_{\Ph}^\mathrm{mod +}$ scenario for $m_{\PA}=150\GeV$ and $\tan\beta=30$ are displayed for comparison.}
\label{fig:mass}

\end{figure*}
The reconstructed jets are required to have transverse momenta $\pt^\text{jet}> 20\GeV$ within the range $\abs{\eta}<2.4$. A multivariate analysis technique is used to remove jets from pileup interactions~\cite{pileup_remove}. Tagging of b quarks in jets relies on the combined secondary-vertex discriminator~\cite{btag_alg}, based on the reconstruction of the secondary vertex from weakly
decaying b hadrons. The discriminant $b_\text{disc}$ is constructed from tracks and secondary vertex information, and helps to distinguish jets containing b, c, or light-flavour hadrons. Jets with an associated $b_\text{disc} > 0.679$ are considered to be b tagged. This value represents a good compromise between efficiency to tag b jets in signal events from AP ($\approx$80\%) and mistagging probability for light-quark jets ($\approx$1\%). Figure~\ref{fig:bdisc} shows the distribution of $b_\text{disc}$ in events that satisfy the selection $\ETm < 35\GeV$,
for the data collected in the two beam energies. For each event, the largest value of $b_\mathrm{disc}$ is selected. The distribution of signal events from the AP process for $m_{\PA}=150\GeV$ and $\tan\beta=30$ is superimposed. Jets originated from b quark fragmentation tend to be emitted more forward in signal events than for $\ttbar$, thus resulting in a lower observed b-jet multiplicity. For this reason the $\ttbar$ background is further suppressed by rejecting events with more than two b-tagged jets, without significantly affecting the selection efficiency for signal.

The events are split into two mutually-exclusive categories. The first category (C1) contains events with at least one jet identified as originated from b-quark fragmentation (b tagged), and provides highest sensitivity to AP production channel. Events that do not contain b-tagged jets are assigned to category 2 (C2), and provide sensitivity to GF production. The dimuon invariant mass distributions for the C1 and C2 categories are shown in Fig.~\ref{fig:mass} for data and simulated events for both centre-of-mass energies. The distributions expected for MSSM Higgs bosons with \mbox{$m_{\PA}=150\GeV$} and $\tan\beta=30$, derived from the $m_{\Ph}^\mathrm{mod +}$ scenario are also given for comparison. A double peak structure around 125 and 150\GeV appears in the C2 category, due to the \Ph boson and \PA{}+\PH bosons, respectively. The lower peak is not visible in C1, as the \Ph production is suppressed in the AP mechanism relative to the GF process.

\begin{figure*}[!htb]
\centering
\includegraphics[width=0.32\textwidth]{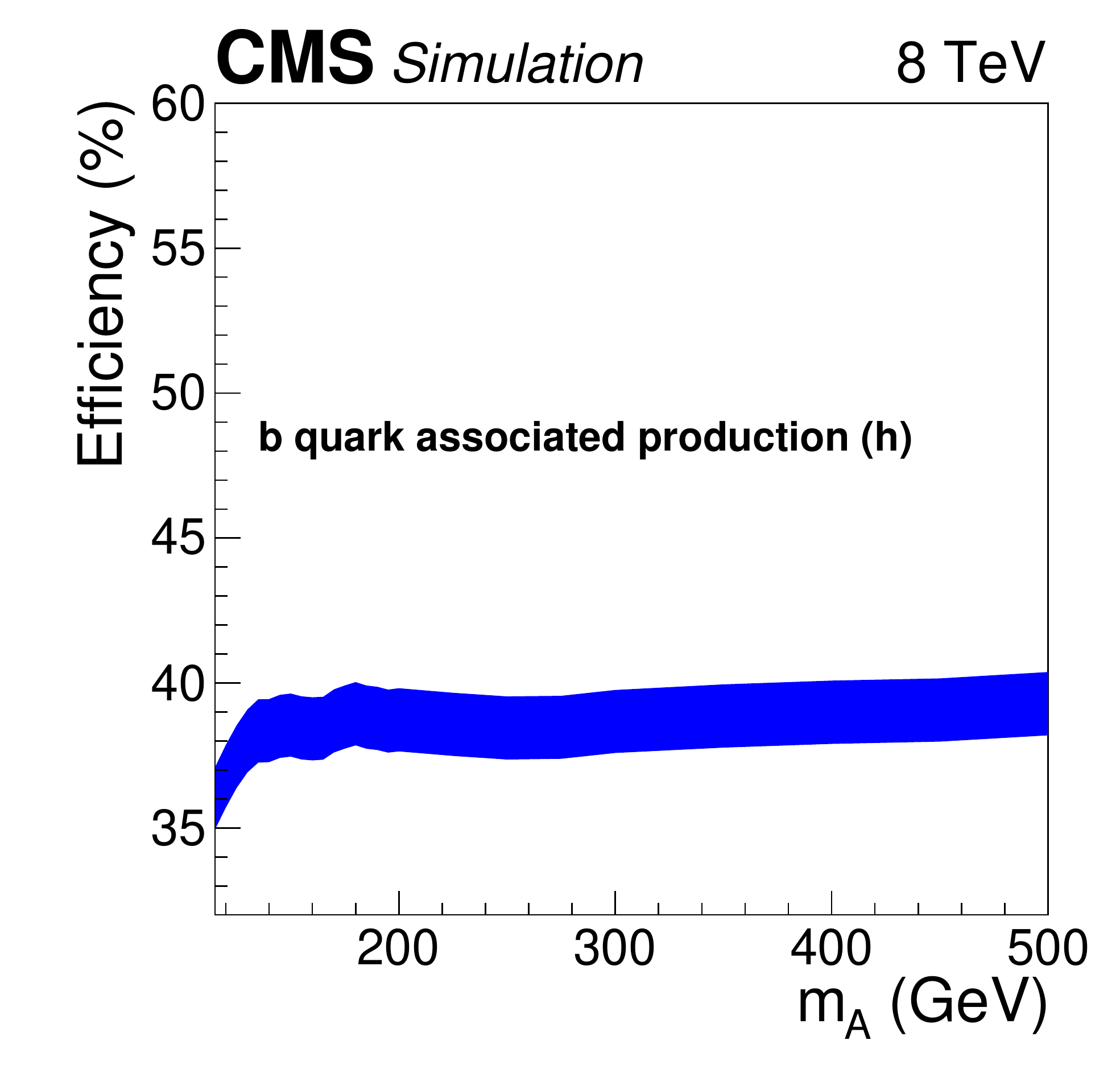}
\includegraphics[width=0.32\textwidth]{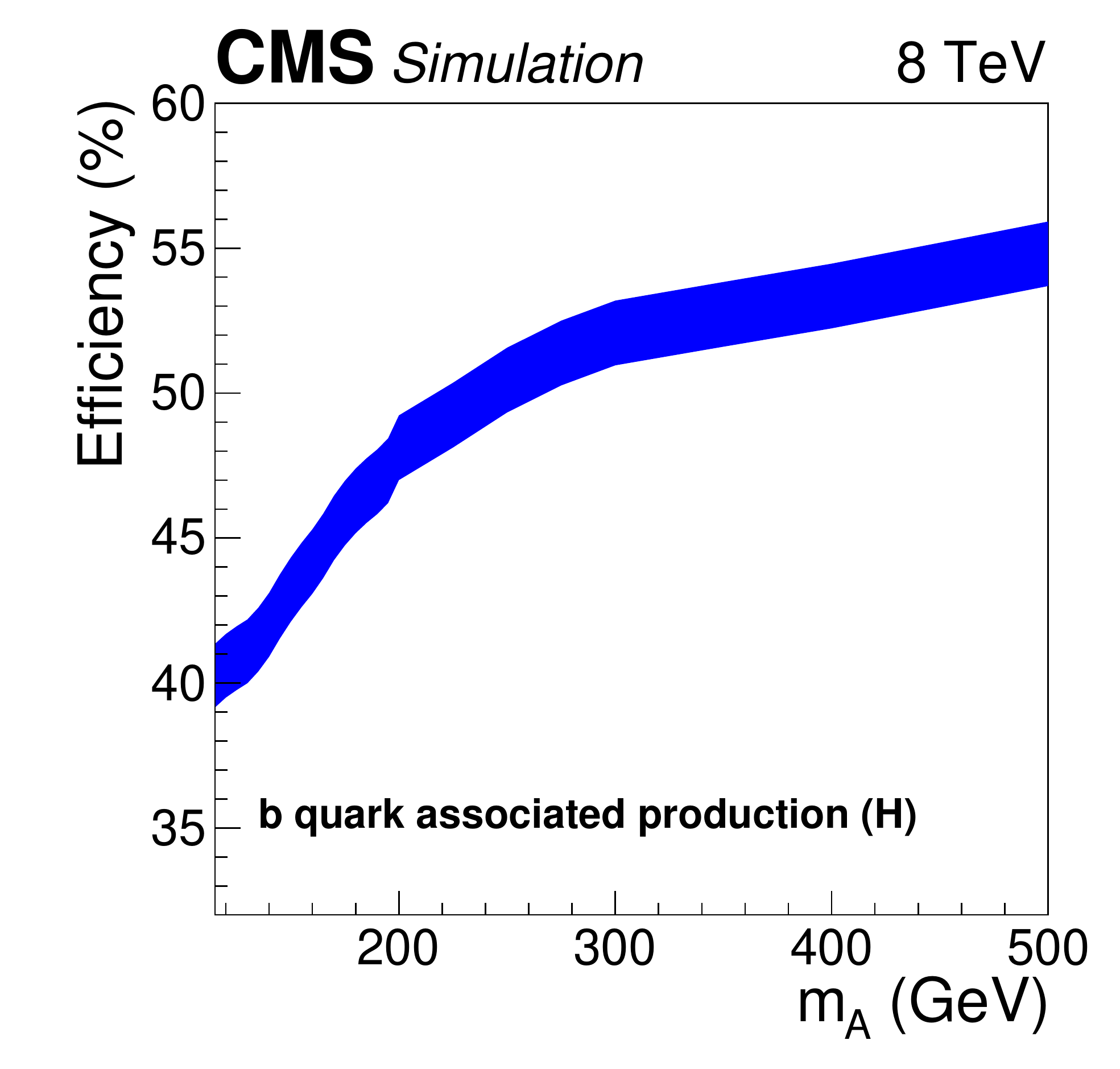}
\includegraphics[width=0.32\textwidth]{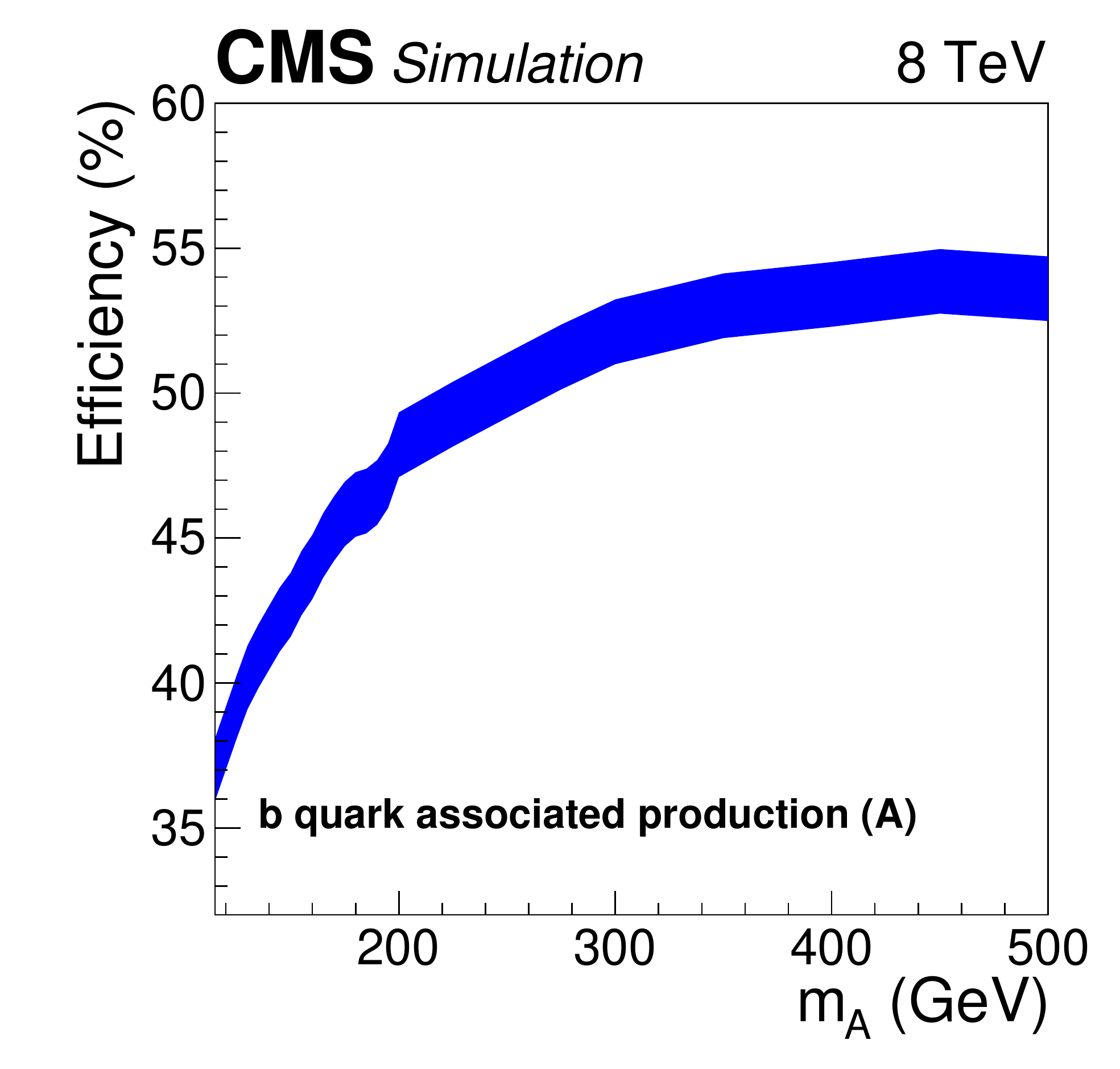}
\includegraphics[width=0.32\textwidth]{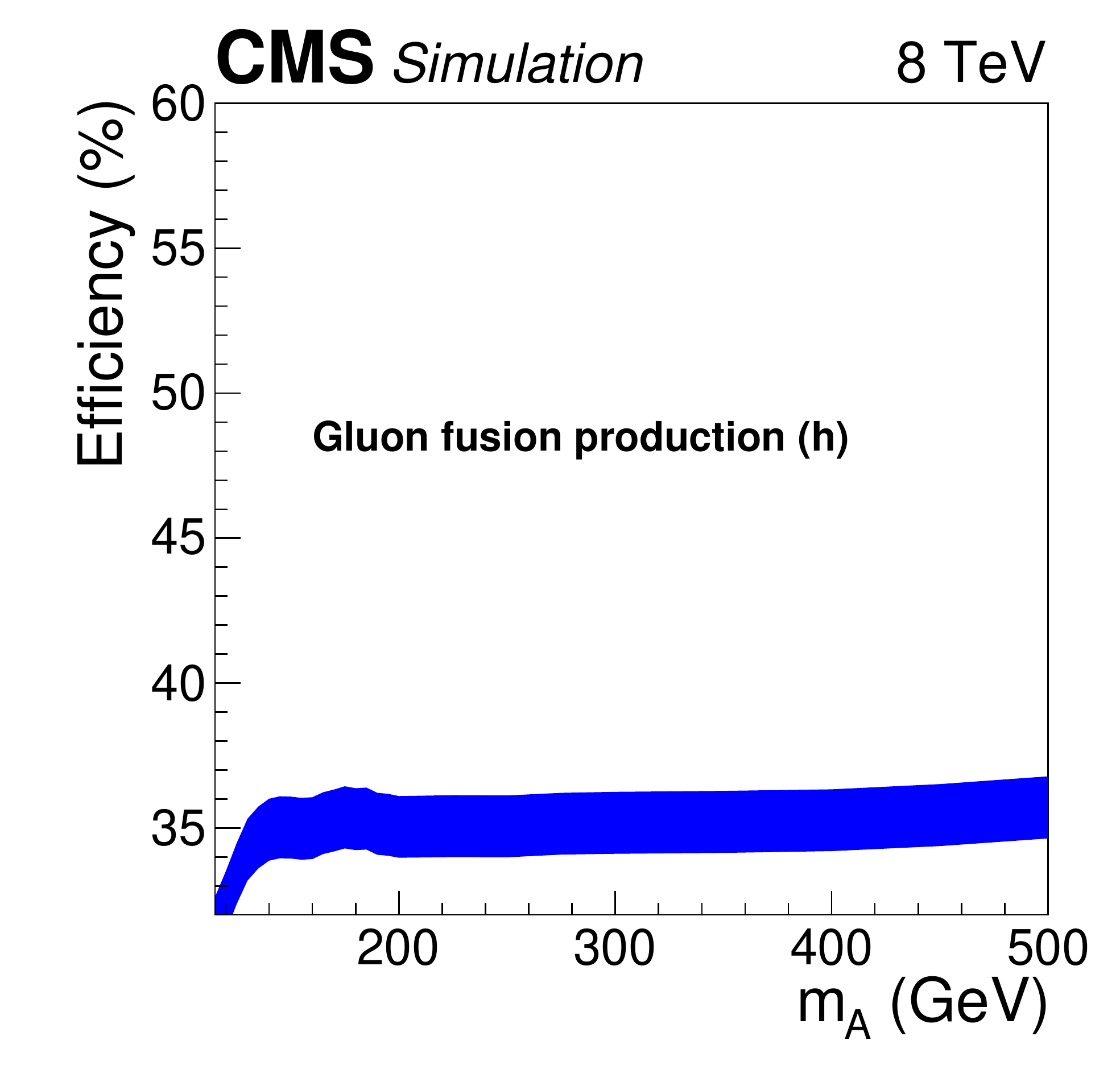}
\includegraphics[width=0.32\textwidth]{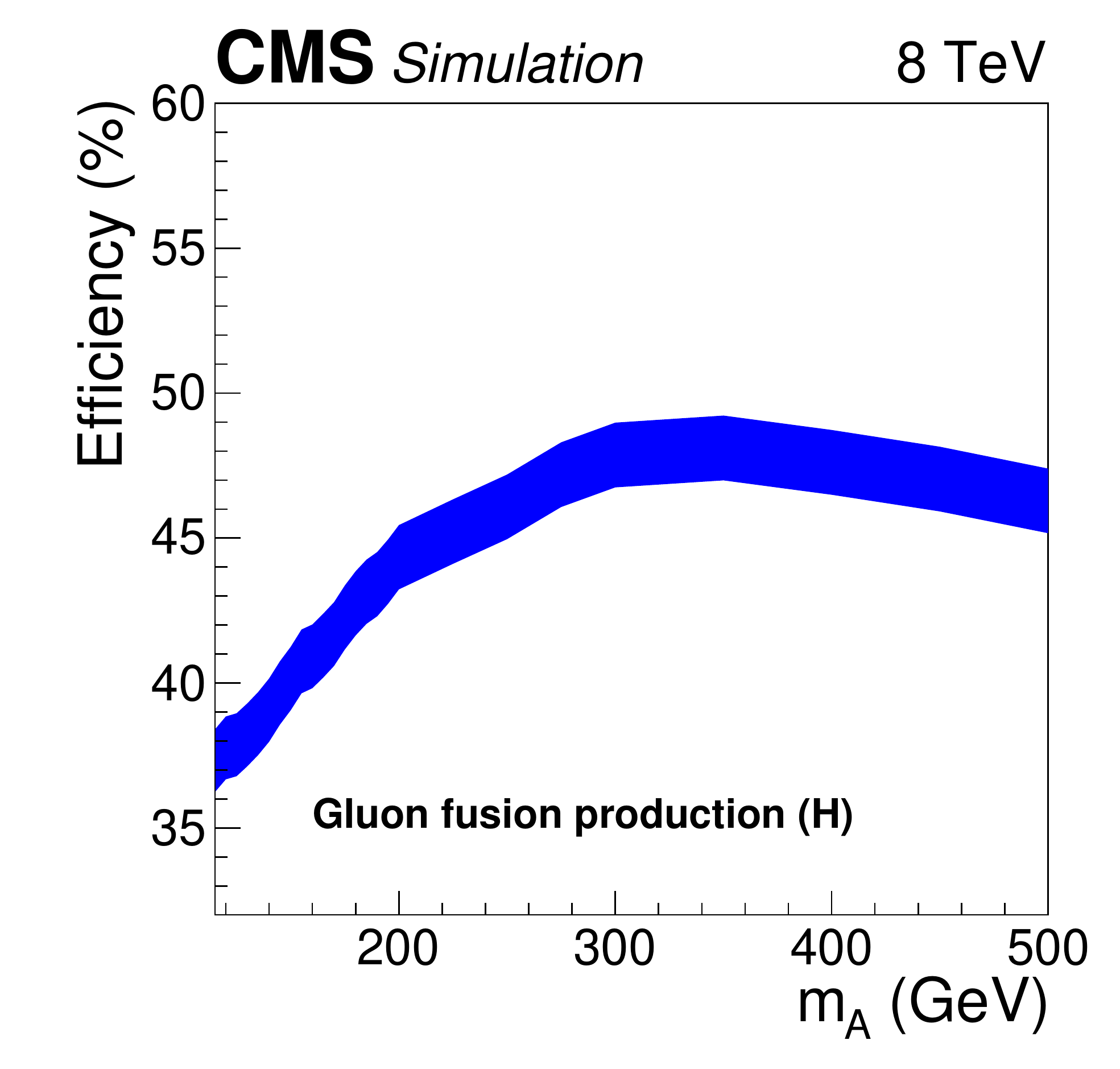}
\includegraphics[width=0.32\textwidth]{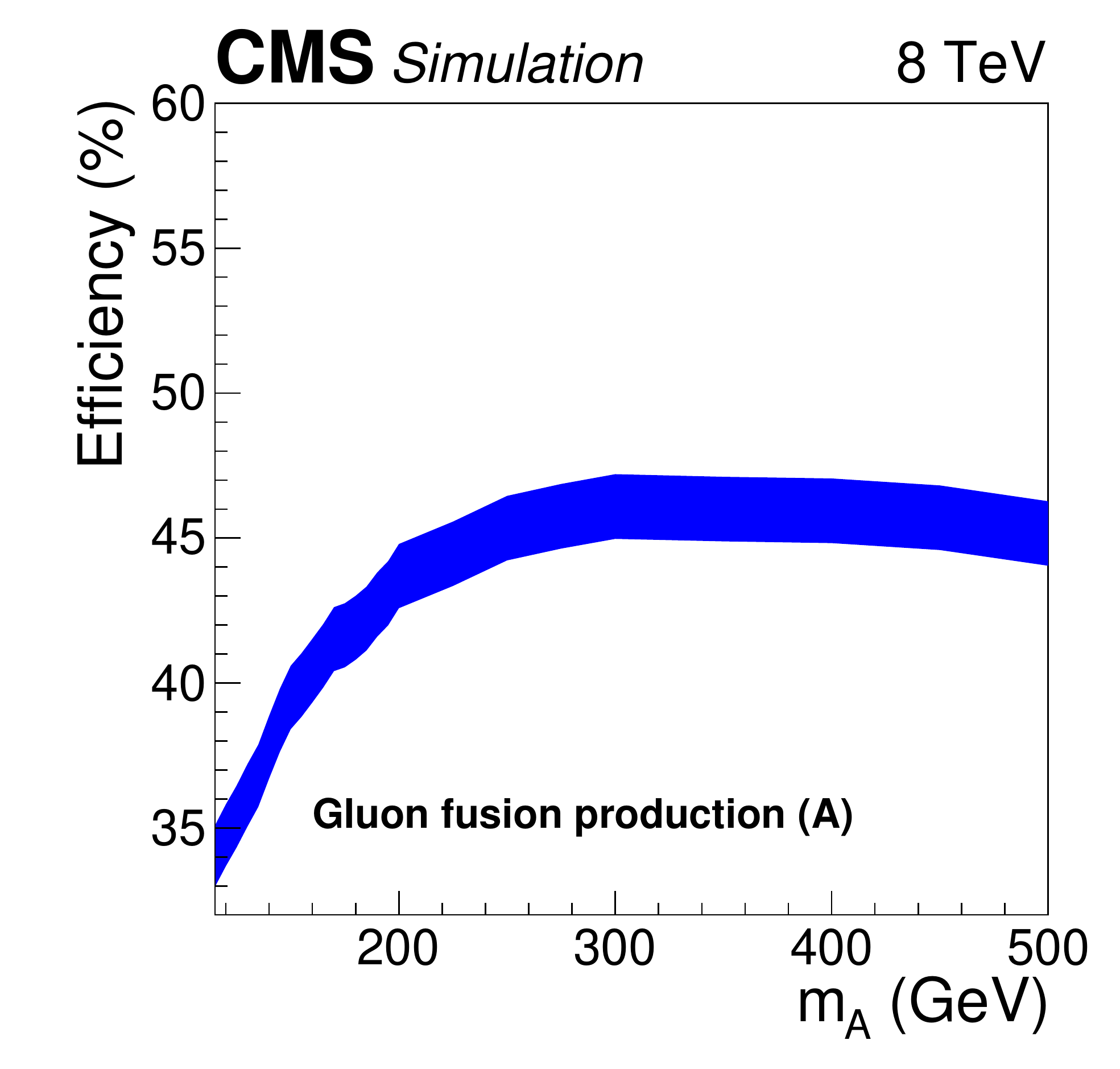}

\caption{Signal efficiency for the AP process at $\sqrt{s}=8\TeV$, shown separately for the three $\phi$ boson
types, (upper left) \Ph, (upper centre) \PH and (upper right) \PA, as a function of $m_{\PA}$. The corresponding efficiency for the GF production process is shown in the lower row. The contributions from the two event categories C1 and C2 are combined. The results are integrated over $\tan\beta$, since the efficiency does not strongly depend on this quantity. The band shows the change in efficiency due to the limited number of simulated events.}
\label{fig:eff_2012}

\end{figure*}

\section{Signal selection efficiency}
\label{sec:Effic}

While the calculations for the MSSM cross sections performed in the narrow-width approximation refer to the on-shell
Higgs boson production, at large values of $m_{\PA}$ and $\tan\beta$ the convolution of the larger intrinsic signal widths with
the parton distribution functions (PDF) results in a non-negligible fraction of signal events produced significantly off-shell.
Events with invariant mass significantly smaller than its nominal value have a lower reconstruction efficiency than
those produced near the mass peak.
For consistency, we define signal efficiency as the probability for a signal event with the generated invariant
mass close to its nominal value to be reconstructed and pass all selection requirements of this analysis. The closeness is defined using
a window of size equal to 3 times the intrinsic signal width (an uncertainty associated with this definition is evaluated using a window
of 5 times its width, as discussed in Section~\ref{sec:sys}). With this definition, the product of the MSSM Higgs boson production cross section, luminosity
and signal
efficiency provides the normalization for the Higgs boson produced near on-shell. The full predicted rate of signal events also
contains an additional off-shell contribution, which varies with $m_{\PA}$ and $\tan\beta$ and is less than 5\% for $m_{\PA}<250\GeV$ and $\tan\beta<15$, and can be as large as 15\% for $m_{\PA}=300\GeV$ and $\tan\beta=30$.

Additional corrections are applied to the signal efficiency to take into account differences between data and simulation in the muon
trigger, reconstruction, and isolation efficiencies. A correction is also applied to account for known data-simulation discrepancies
in the b tagging efficiency and mistagging probability. The corrections are summarized by a weight factor,
which is assigned to each signal event.
The average of the weight factors computed over all the events is very close
to one, reflecting the fact that the simulation describes the data with good accuracy.

Figure~\ref{fig:eff_2012} shows the signal efficiency at $\sqrt{s}=8\TeV$ for AP (top) and GF (bottom) process after combining the two event categories C1 and C2. The efficiencies at $\sqrt{s}=7\TeV$ are similar. The band in the figure
represents the variation of the efficiency due to the limited statistics of the samples used.
The relative amount of AP and GF events in the two event categories varies with $m_{\PA}$ and $\tan\beta$, since the production cross sections of the two processes depend on these parameters.
For example, in the case $m_{\PA}=150\GeV$
and $\tan\beta=30$, more than 90\% of the signal events in C1 would be from AP production,
and about 60\% in C2. For $m_{\PA}=150\GeV$ and $\tan\beta=5$, where the GF contribution becomes more relevant,
the content of AP events would be 60\% in C1 and only 15\% in C2.

\section{Fit procedure}
\label{sec:Fit}
\begin{figure}[!ht]
\centering
\includegraphics[width=0.48\textwidth]{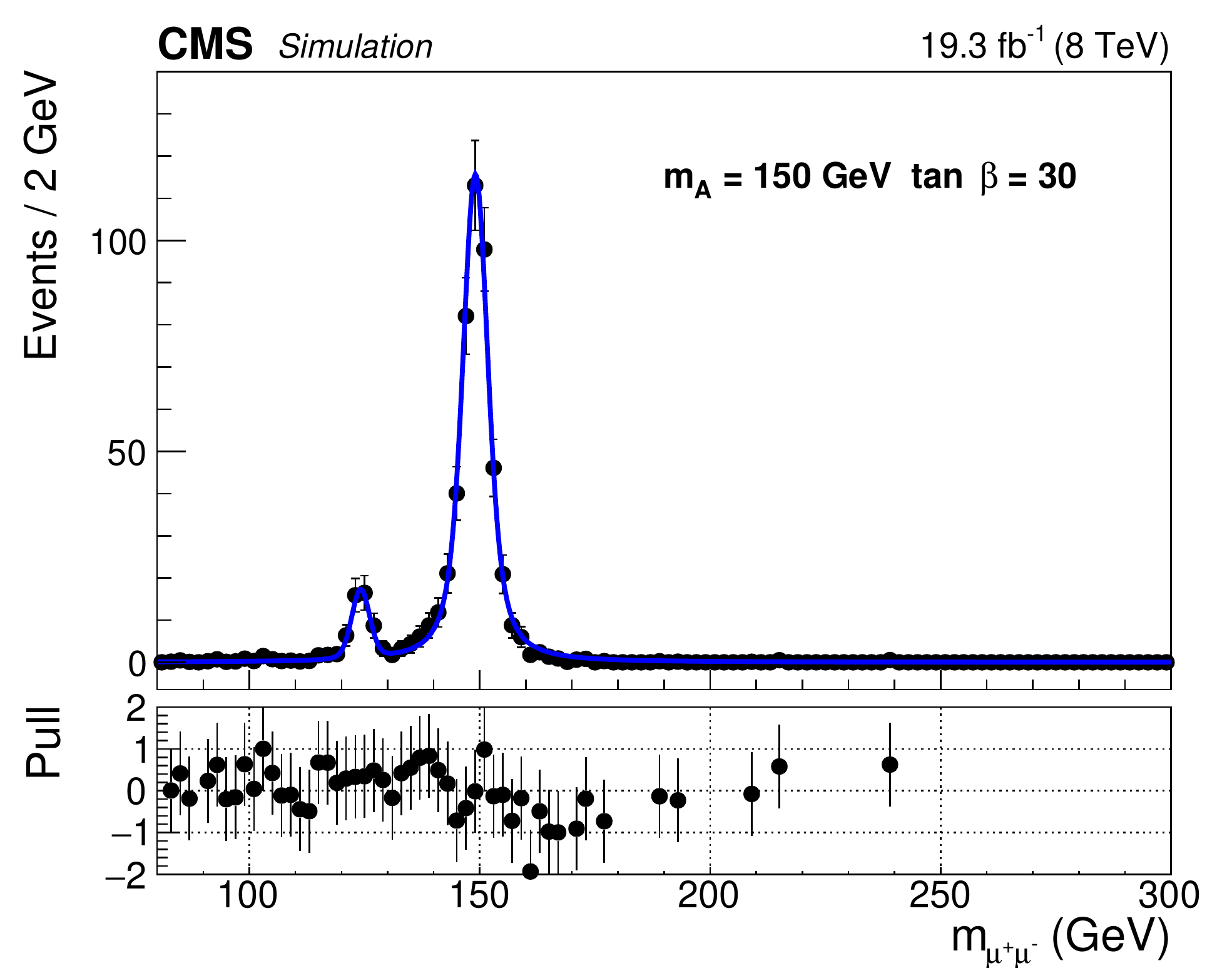}\label{fig:signalfit}
\includegraphics[width=0.48\textwidth]{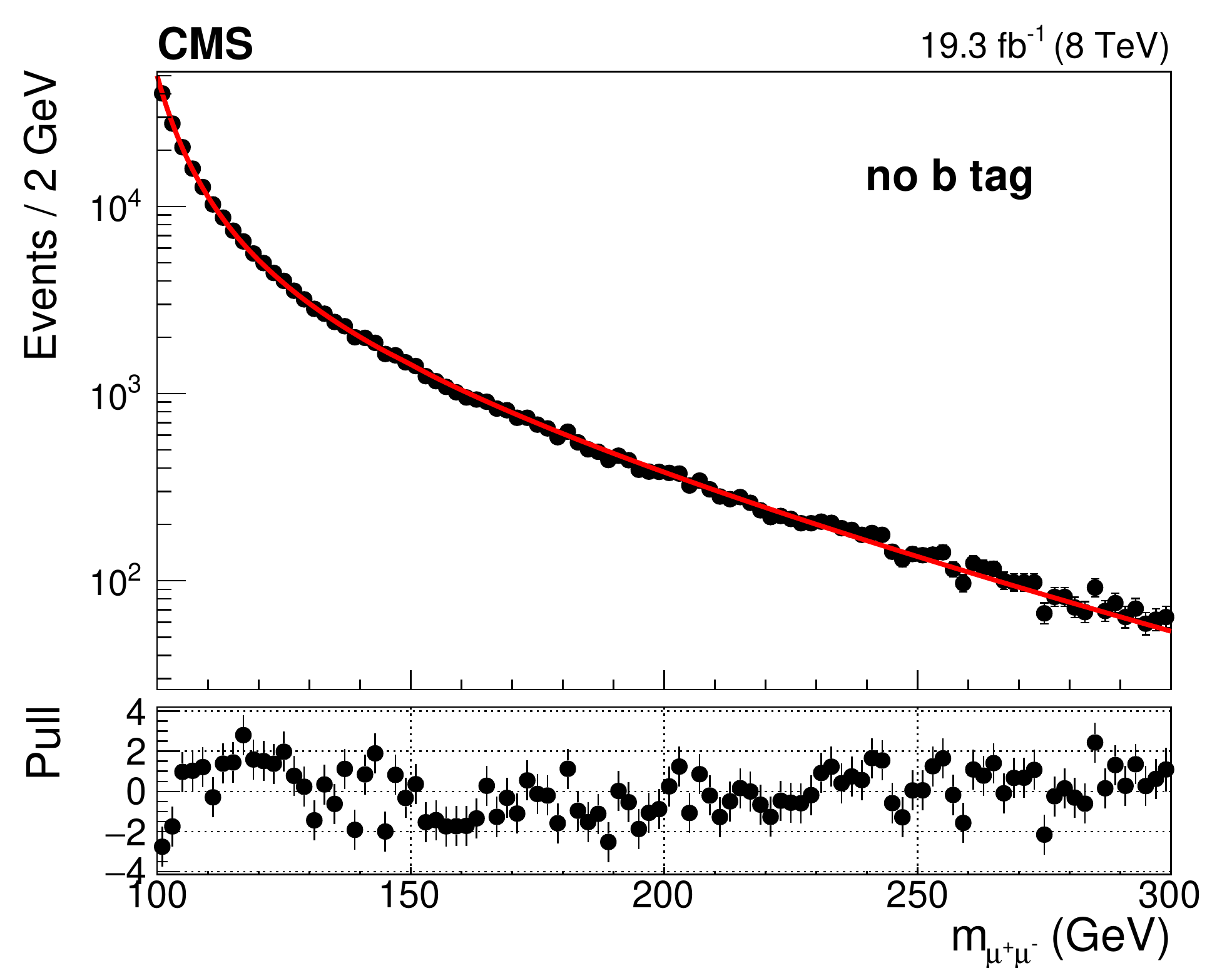}\label{fig:FfitCat2}
\caption{Invariant mass distribution of the expected signal
for $m_{\PA}= 150\GeV$
and $\tan\beta = 30$ (\cmsLeft),
and an example of the fit to
the data at $\sqrt{s}=8\TeV$ including the same signal assumption (\cmsRight). The distribution represents the expected number of events for an integrated luminosity of 19.3\fbinv. For each plot
the pull of the fit as a function of the dimuon invariant mass is shown.}
\label{fig:fit}

\end{figure}

The procedure described below is applied separately to C1 and C2 events.
The event selection criteria are applied to the simulated samples listed in Table~\ref{tab:MC_signal}. For each sample, and for each of the three $\phi$ bosons,
the invariant mass distribution of the events that pass the event selection is approximated with a
Breit--Wigner function convolved with a Gaussian, that accounts for detector resolution. This analytical expression provides a good description of the signal shape for all the $m_{\PA}$ and $\tan\beta$ values. The three functions are denoted $F_{\Ph}$, $F_{\PH}$, and $F_{\PA}$, and contain the mass and width of the Breit--Wigner and the width of the Gaussian as free parameters.
The function $F_\text{sig}$ represents the expected signal yield, and it is a linear combination of the three functions described above:
\begin{equation}
F_\text{sig} = w_{\Ph} \, F_{\Ph} + w_{\PH} \, F_{\PH} +  w_{\PA} \, F_{\PA},
\label{eq:signalfit}
\end{equation}
where $w_{\Ph}$, $w_{\PH}$, and $w_{\PA}$, are the number of events containing \Ph, \PH, and \PA bosons, respectively, calculated according to their expected production cross sections. An example of this procedure is shown in Fig.~\ref{fig:fit}\,(\cmsLeft) for $m_{\PA}= 150\GeV$ and $\tan\beta = 30$. The highest peak represents the superposition of the contributions from \PH and \PA bosons, that in this case are almost degenerate in mass.

Since the Drell--Yan muon pair production is the dominant background process, it is modeled by a Breit--Wigner function
plus a photon-exchange term,
which is proportional to $1/m_{\PGmp\PGmm}^{2}$. Defining $m=m_{\PGmp\PGmm}$, the function $F_\text{bkg}$ becomes:
\begin{equation}
F_\text{bkg} =  \re^{\lambda m} \left[\frac{f_{\Z}}{N^\text{norm}_{1}} \, \frac{1}{{(m-m_{\Z})}^2 +\frac{\Gamma^2_{\Z}}{4}} + \frac{(1-f_{\Z})}{N^\text{norm}_{2}} \, \frac{1}{{m}^2} \right],
\label{eq:bkg_function}
\end{equation}
where $e^{\lambda m}$ describes the effects of the PDF, and the $N^\text{norm}_i$ terms correspond to the integral of the corresponding functions in the chosen mass range. The quantity $f_{\Z}$ represents the contribution of the Breit--Wigner term relative to the photon-exchange term. The quantities $\lambda$ and $f_{\Z}$ are free parameters of the fit. The parameters $\Gamma_{\Z}$ and $m_{\Z}$ are determined separately for the C1 and the C2 events from a fit to the $m_{\PGmp\PGmm}$ distribution in the mass range  of the \Z boson between 80 and 120\GeV.
The fit provides the effective values of such quantities, that include detector and resolution effects for each set of data. Their values are used in $F_\text{bkg}$ and are kept constant in the fit.

A linear combination of the two functions for the expected signal and background is then used in an unbinned likelihood fit to the data:
\begin{equation}
F_\text{fit} = (1 - f_\text{bkg}) \, F_\text{sig} + f_\text{bkg} \, F_\text{bkg}.
\label{eq:Ffit}
\end{equation}

The parameters that describe the signal are determined in the fit of the simulated signal to Eq.~(\ref{eq:signalfit}), for each pair of $m_{\PA}$ and $\tan\beta$ values. Subsequently, they are fixed in $F_\text{fit}$, where the free parameters are the quantities $\lambda$, $f_{\Z}$, and $f_\text{bkg}$.
The fraction of signal events is defined as $f_\text{sig}=(1-f_\text{bkg})$. The data are fitted to $F_\text{fit} $ in the mass range from 115 to 300\GeV for each point in the $m_{\PA}$ and $\tan\beta$ parameter space.
As an example, the fit to the data of C2 at $\sqrt{s}=8\TeV$ is illustrated in Fig.~\ref{fig:fit}\,(\cmsRight),
assuming a signal with $m_{\PA}= 150\GeV$ and $\tan\beta = 30$.

\section{Systematic uncertainties}
\label{sec:sys}
The following sources of systematic uncertainties are taken into account, and the impact of one standard deviation change is reported in terms of a variation in the nominal signal efficiency defined in Section~\ref{sec:Effic}.

The limited number of simulated events introduces an uncertainty in the signal selection efficiency that is at most 2.0\%. The muon trigger, reconstruction, identification, and isolation efficiencies are determined from data using a tag-and-probe technique~\cite{MuonPOG_efficSys}.
The uncertainty in the trigger efficiency correction is 0.5\%, whereas 1.0\% is assigned to the combination of uncertainties in muon reconstruction and identification, as well as on isolation efficiencies.

A systematic uncertainty in the pileup multiplicity is evaluated by changing
the total cross section for inelastic pp collisions in simulation. The corresponding uncertainty on the signal
efficiency is at most 0.8\% in both categories.

The event fractions in the two categories depend on the b tagging efficiency and the mistagging probability.
The uncertainty in the b tagging efficiency is estimated by comparing data and simulated events with samples of
enriched b quark content and different topologies, as described in Ref.~\cite{btag_alg}. The uncertainty in the efficiency to detect b jets is about 3.0\%. Similarly, the uncertainty in the mistagging rate is about 10\%. Their overall contribution to the selection efficiency is
weighted by the fraction of AP and GF events that are expected in each event category, which depends on $m_{\PA}$ and $\tan\beta$. The largest overall uncertainty is 3.0\% for C1, and 0.4\% for C2 events.

The jet energy scale uncertainty is estimated by smearing the
jet momentum by a factor depending on $\pt$ and $\eta$ of each jet,
as described in Ref.~\cite{jes}. The effect on signal selection efficiency
is 4.0\% for events that belong to the C1 and 0.5\% for the
C2 categories, at $\sqrt{s}=8\TeV$. For $\sqrt{s}=7\TeV$ the corresponding
numbers are 3.8\% and 0.6\%. The uncertainty in the $\ETm$ scale and resolution is estimated through comparisons between data and simulation~\cite{METsys,METsys2}. The effect on the signal selection efficiency is 3.0\% and 2.0\%, the same for both categories, for the sample with $\sqrt{s}= 8$ and 7\TeV, respectively. The uncertainty in the integrated luminosity
is 2.6\% and 2.2\% at $\sqrt{s}=8$ and 7\TeV, respectively~\cite{lumifinal,lumifinal2}.

Uncertainties due to the choice of PDF set affect the signal efficiency, and are studied using the PDF4LHC~\cite{PDF4LHC} prescription.
The renormalization and factorization scales in the calculations and their changes are
summarized in Refs.~\cite{cs_higgs1,cs_higgs2,cs_higgs3}. The effect on the signal selection efficiency varies from 1.0\% to 3.0\% over the $m_{\PA}$ and $\tan\beta$ parameter space. The choice of 3.0\% is taken as the systematic uncertainty.

The efficiency is determined for events with generated mass values within a window of a factor of 3 of the intrinsic width of the Higgs boson, as described in Section~\ref{sec:Effic}. The difference relative to the efficiency obtained using a cutoff of a factor of 5 of the intrinsic width is assigned as a systematic uncertainty. The uncertainty is between 1\% to 3\% for the C1 and 1\% to 5\% for the C2 categories.

Table~\ref{tab:sys} lists the systematic uncertainties that affect the determination of signal efficiency. The impact of these systematic uncertainties on the exclusion limits that will be presented in Section~\ref{sec:Limits} is negligible compared to the statistical uncertainty. All the systematic uncertainties in Table~\ref{tab:sys} are correlated for the $\sqrt{s}=7\TeV$ and 8\TeV data, with the exception of the uncertainties related to the limited MC statistics and the integrated luminosity.

The uncertainties in the MSSM cross sections
depend on $m_{\PA}$, $\tan\beta$, and the scenario, and are provided by the LHC Higgs Cross Section Working Group~\cite{cs_higgs1,cs_higgs2,cs_higgs3}. The signal events are generated using \PYTHIA, assuming the parameters of the $m_{\Ph}^\text{max}$ scenario, as discussed in Section~\ref{sec:Data}. The different benchmarks are expected to affect the production cross section, but not the kinematic properties of the events related to Higgs boson production and decay.
To check this assumption, events are generated with \PYTHIA using the parameters for the $m_{\Ph}^\mathrm{mod +}$, $m_{\Ph}^\mathrm{mod -}$, light stop and light stau benchmarks, assuming $m_{\PA}=150\GeV$ and $\tan\beta=20$. The events are generated for both the GF and the AP mechanisms,
and the Higgs boson $\pt$ and the $\ETm$ of the events are compared at generator level for the various benchmark scenarios.
No significant differences are observed in the distributions of these quantities.

\begin{table}[!ht]
\topcaption{Sources of systematic uncertainties for C1 and C2 event categories that affect the signal efficiency at $\sqrt{s}=8\TeV$. They are expressed in terms of relative signal selection efficiency. When the systematic uncertainty at $\sqrt{s}=7\TeV$ differs from $\sqrt{s}=8\TeV$,
the corresponding value is quoted in parenthesis.}
\centering
\begin{tabular}{  l cc}
\hline
\multirow{2}{*}{Source}                  & \multicolumn{2} {c} {Systematic uncertainty (\%)} \\
                        & C1     & C2                               \\ \hline
MC statistics           & 2.0       & 2.0                                 \\
Trigger efficiency      & 0.5       & 0.5                                 \\
Muon efficiency         & 1.0       & 1.0                                 \\
Muon isolation          & 1.0       & 1.0                                 \\
Pileup                 & 0.8       & 0.8                                 \\
b tagging                   & 3.0       & 0.4                                 \\
Jet energy scale        & 4.0 (3.8) & 0.5 (0.6)                           \\
$\ETm$  & 3.0 (2.0) & 3.0 (2.0)                           \\
Integrated luminosity       & 2.6 (2.2) & 2.6 (2.2)                           \\
PDFs                  & 3.0      & 3.0                                \\
Width correction        & 1--3 & 1--5 \\
\hline

\end{tabular}

\label{tab:sys}
\end{table}

Since the number of background events is determined through a fit to the data,
an additional systematic uncertainty arises from the possibility
that the background parametrization may not adequately describe the data as a function of the
dimuon invariant mass.  A method similar to that described in Ref.~\cite{newboson2} is used to evaluate the effect, by estimating the uncertainty through the bias in
terms of the number of signal events that are found when fitting the signal + background model (as described in Section~\ref{sec:Fit}) to pseudo-data generated for different alternative background models. Such alternative background parametrizations include Bernstein polynomials and combinations of Voigtian and exponential functions. Bias estimates are performed for mass points between $m_{\PA} =115$ and 300\GeV. For each $m_{\PA}$ value, the largest bias among the tested functions is taken as the resulting uncertainty. The bias is implemented as a floating additive contribution to the number of signal events,
constrained by a Gaussian probability density with mean of zero and width set
to the systematic uncertainty. The width of the Gaussian is the largest systematic uncertainty, and the effect is to increase the expected limit on the presence of a signal by 20\% in the region near $m_{\PA}=120\GeV$ and by about 10\% at larger mass values.

In the mass range between 115 and 300\GeV, that is relevant for this analysis, the mass resolution is estimated to be between 1.2 and 4\GeV.
Uncertainties in the muon momentum determination can affect the invariant mass measurement, and have been carefully studied in data and simulation~\cite{MuonPOG_efficSys}. The dimuon invariant mass resolution for masses above the \Z peak has been previously studied in the search for a SM Higgs decaying to a dimuon pair~\cite{Htomu_CMS}. The mass resolution determined from data at the \Z mass value is 1\GeV, in excellent agreement with the prediction from simulation. This value is consistent with the mass resolution of 1.2\GeV that we estimate from simulation for a mass of 115\GeV, that corresponds to the lower edge of the Higgs mass range considered in this analysis.

The overall capability of the analysis to detect the presence of a signal is verified by introducing a hypothetical simulated signal in the data using the shape parametrization discussed in Section~\ref{sec:Fit}.
The average measured number of signal events is found to be within 1.3\% of the injected signal for the C1 category,
and within 4.3\% for the C2 category. These differences are assigned as systematic uncertainties.

\section{Results}
\label{sec:Limits}
\begin{figure}[!ht]
\centering
\includegraphics[width=\cmsFigWidth]{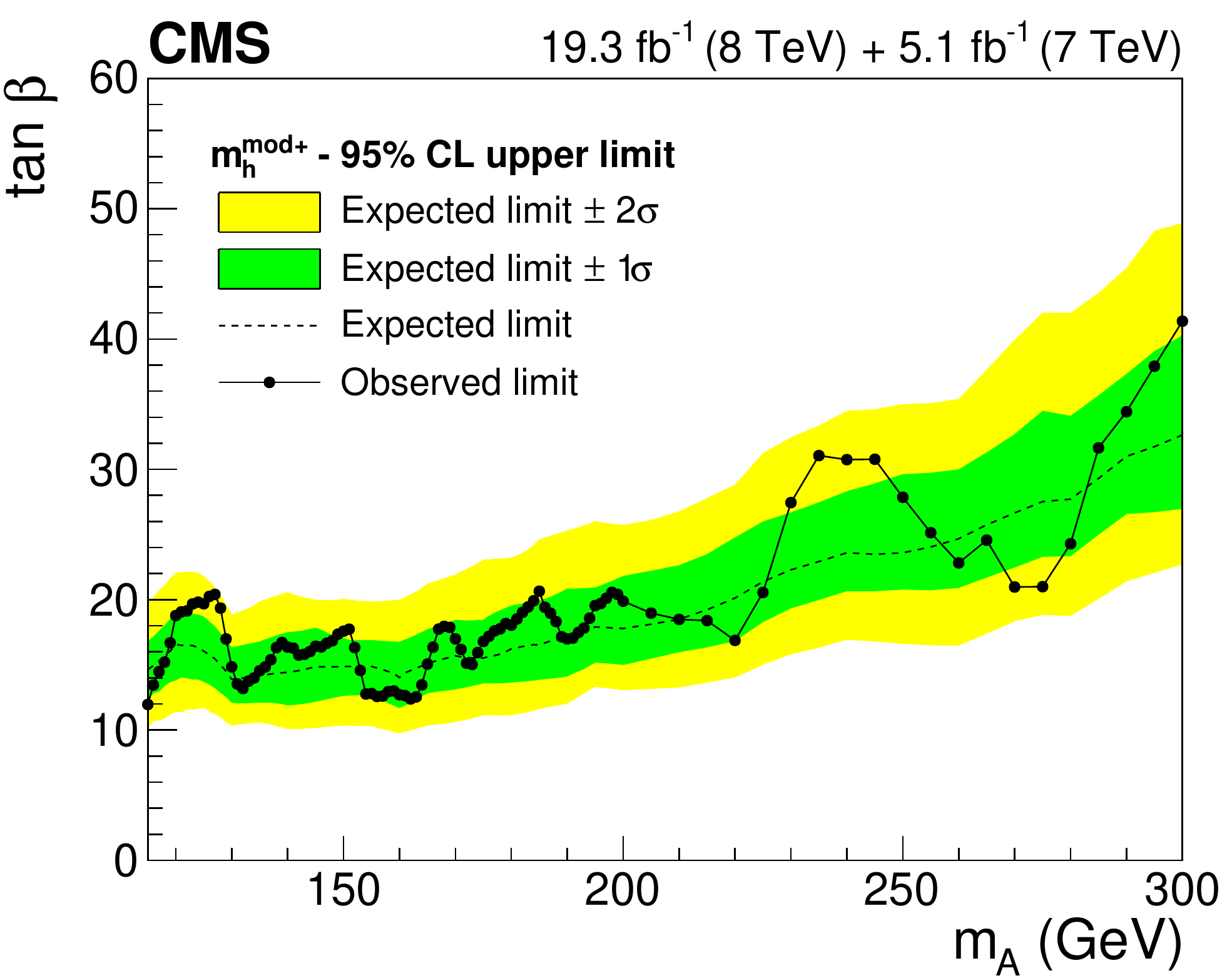}
\caption{The 95\% CL upper limit on $\tan\beta$ as a function of $m_{\PA}$, after combining the data from the two event categories at the two centre-of-mass energies (7 and 8\TeV). The results are obtained in the framework of the $m_{\Ph}^\mathrm{mod +}$ benchmark scenario. }
\label{fig:Combination}

\end{figure}

No evidence of MSSM Higgs bosons production is observed in the mass range between 115 and 300\GeV, where the analysis has been performed. Upper limits at 95\% confidence level (CL) on the parameter $\tan\beta$ are computed
using the CL$_\mathrm{s}$ method~\cite{Read,Junk}, which is a modified frequentist criterion, and are presented as a function of $m_{\PA}$. Systematic uncertainties are incorporated as nuisance parameters
and treated according to the frequentist paradigm~\cite{HiggsCombination}. The results are obtained from a combination of both event categories and centre-of-mass energies. For each value of $m_{\PA}$, the value of $\tan\beta$ at which the CL exceeds 95\% is chosen to define the exclusion limit on that $m_{\PA}$. This is performed for all the $m_{\PA}$ values and the results are shown in Fig.~\ref{fig:Combination}. These results are obtained within the $m_{\Ph}^\mathrm{mod +}$ scenario. The observed upper limits range from $\tan\beta$ of about 15 in the low-$m_{\PA}$ region, to above 40 at $m_{\PA}=300\GeV$. For larger values of $m_{\PA}$ the uncertainty on the $\tan\beta$ upper limit becomes large, exceeding $\tan\beta=50$, for which the MSSM cross-section predictions are not reliable.

\begin{figure}[!ht]
\centering
\subfigure{\includegraphics[width=\cmsFigWidth]{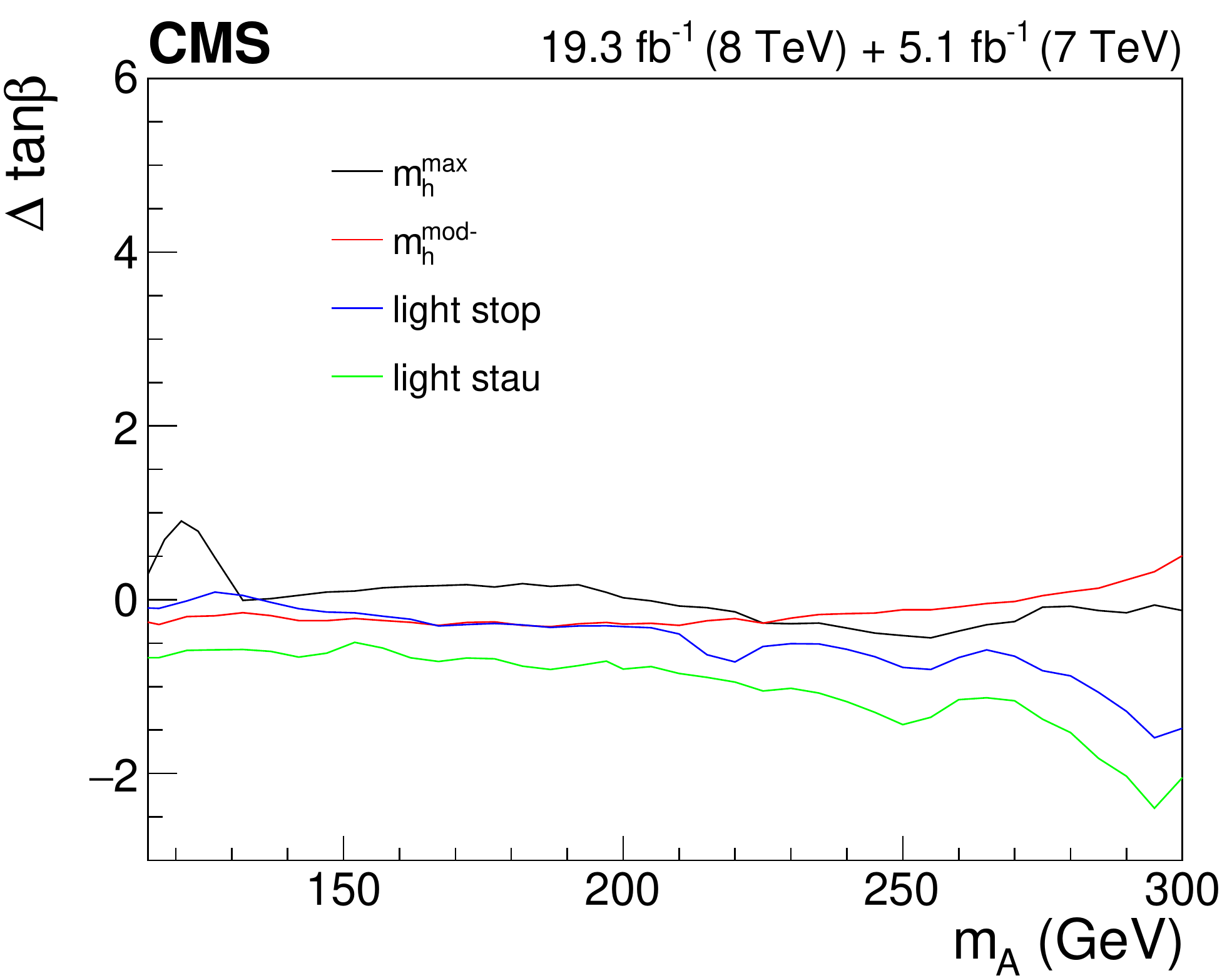}}
\caption{Comparison of the 95\% CL exclusion limits on $\tan\beta$ obtained within MSSM benchmark models, as a function of
$m_{\PA}$. The quantity \mbox{$\Delta \tan\beta = \tan\beta_{m_{\Ph}^\mathrm{mod +}} - \tan\beta_\text{scenario}$} represents the difference in $\tan \beta$ at which the 95\% CL limit is obtained for alternative scenarios.
 }

\label{fig:compare_limits}

\end{figure}

A comparison with the results obtained for the $m_{\Ph}^\mathrm{mod -}$, $m_{\Ph}^\text{max}$, light stop and light stau scenarios is also performed. The exclusion limits computed within these other benchmark models are all very similar. For any value of $m_{\PA}$, the quantity $\Delta \tan\beta = \tan\beta_{m_{\Ph}^\mathrm{mod +}} - \tan\beta_\text{scenario}$ represents the difference of the
$\tan \beta$ values at which the 95\% CL limit is determined if an alternative scenario is used. Figure~\ref{fig:compare_limits} shows the quantity $\Delta \tan\beta$ as a function of $m_{\PA}$ for all the tested scenarios. For most $m_{\PA}$ values, the 95\% CL limits on $\tan \beta$ computed within a given scenario differ by less than one unit from the results obtained within the $m_{\Ph}^\mathrm{mod +}$ scenario.

\begin{figure}[!ht]
\centering
\includegraphics[width=0.48\textwidth]{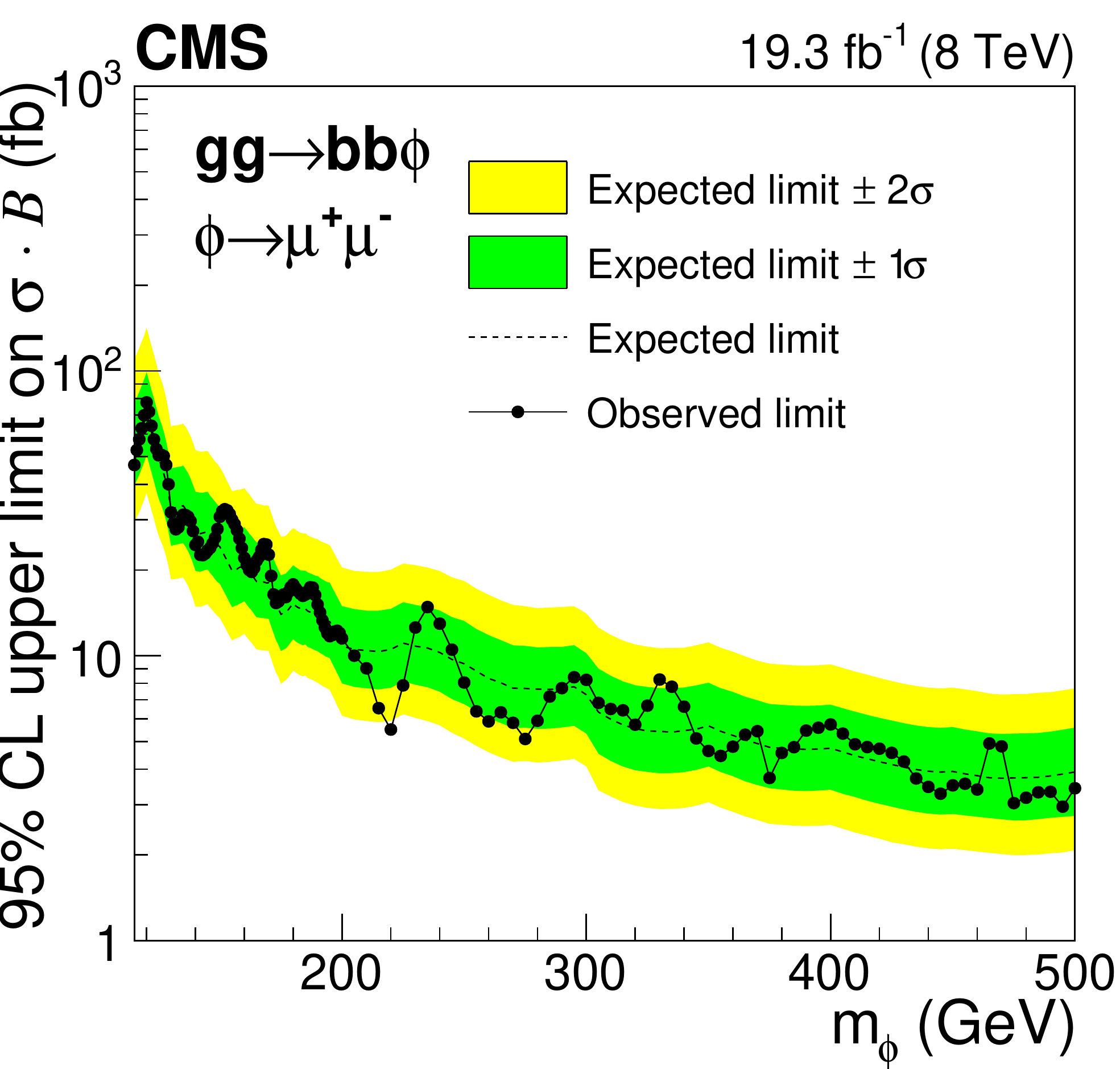}
\includegraphics[width=0.48\textwidth]{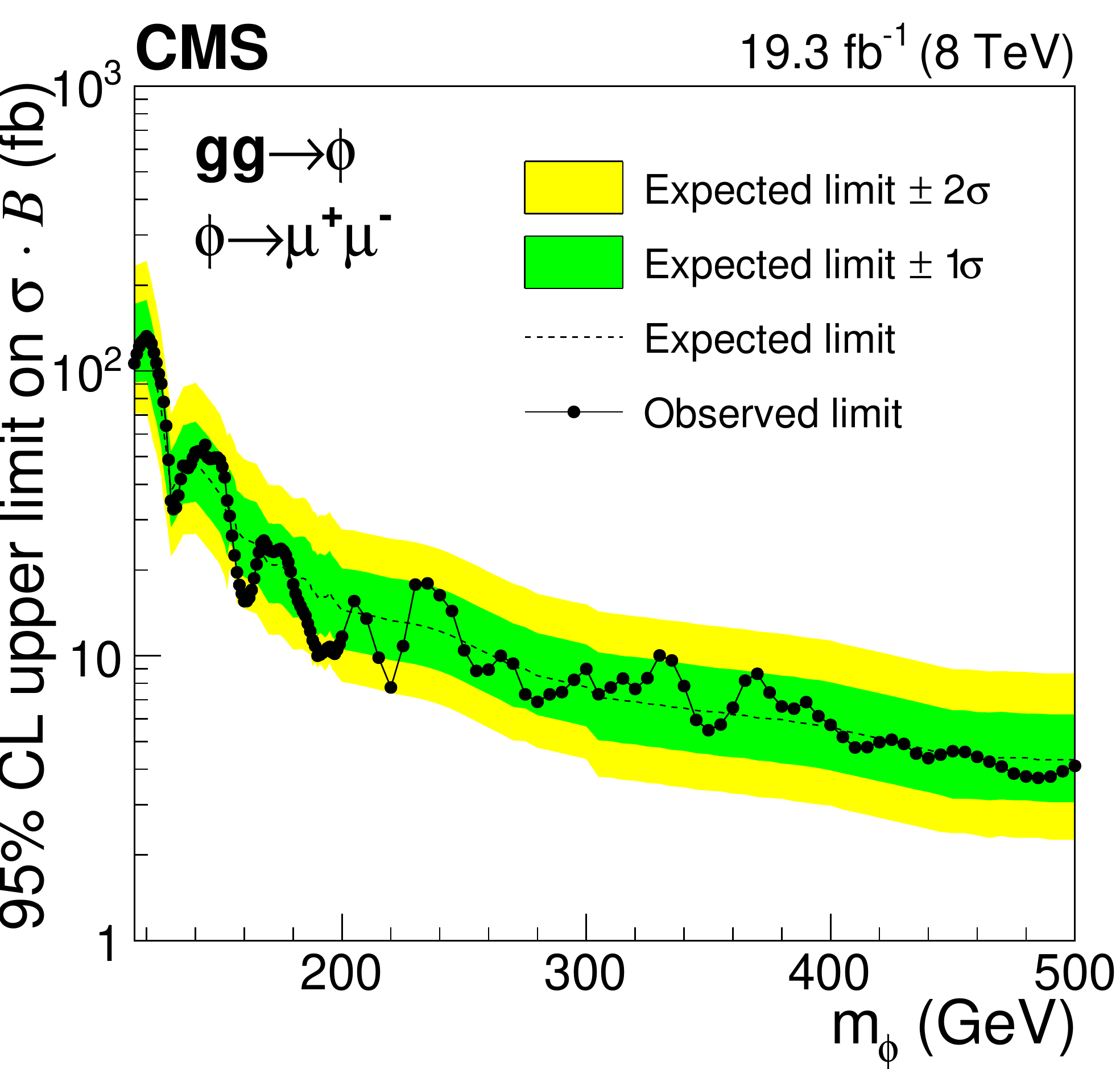}
\caption{The 95\% CL limit on the product of the cross section and the decay branching fraction to two muons as a function of $m_{\phi}$,
obtained from a model independent analysis of the data.  The results refer to (\cmsLeft) b quark associated and (\cmsRight) gluon fusion production, obtained using data collected at $\sqrt{s}=8\TeV$. }
\label{fig:ModelIndep}

\end{figure}

Limits on the production cross section times decay branching fraction $\sigma \, \mathcal{B} ({\phi} \to \PGmp\PGmm) $
for a generic single neutral boson $\phi$ are determined. In this model independent analysis no assumption is made on the cross
section, mass, and width of the $\phi$ bosons, which is sought as a single resonance with mass $m_{\phi}$.
The analysis is performed assuming the narrow width approximation, for which the intrinsic width of the signal
is smaller than the invariant mass resolution. For this purpose the simulated signal of the A boson for the case $\tan\beta$ = 10 is
used as a template to compute the detection efficiency for a generic $\phi$ boson decaying to a muon pair.
The single $\phi$ boson is assumed to be produced entirely either
via the AP or the GF process, and the search for a single resonance with mass $m_{\phi}$ is performed.
The 95\% CL exclusion on $\sigma \, \mathcal{B}({\phi} \to \PGmp\PGmm) $  is determined as a function of $m_{\phi}$,
separately for the two production mechanisms. The combination of events belonging to C1 and C2 is shown in Fig.~\ref{fig:ModelIndep},
assuming the $\phi$ boson is produced either via the AP or the GF process. Only data collected at $\sqrt{s}=8\TeV$ are used, as they
provide a better sensitivity because of the higher luminosity. In addition, since the $\phi$ production cross section depends
on the centre-of-mass energy, a combination with the 7\TeV results would introduce a model dependence in the description of
the cross section evolution with energy.

\section{Summary}
A search has been performed for neutral MSSM Higgs bosons decaying to $\PGmp\PGmm$ from pp collisions collected with the CMS
experiment at $\sqrt{s}=7$ and 8\TeV, corresponding to integrated luminosities of 5.1 and 19.3\fbinv, respectively. The analysis
is sensitive to Higgs boson production via gluon fusion, and via association with a $\bbbar$ quark pair. The
results of the search, which has been performed in the mass range between 115 and 300\GeV, are
presented in the $m_{\Ph}^\mathrm{mod +}$ framework of the MSSM. With no evidence for MSSM Higgs boson production, this analysis excludes at 95\% CL values of $\tan\beta$ larger than
40 for Higgs boson masses up to 300\GeV. Comparisons with $m_{\Ph}^\mathrm{mod- }$, $m_{\Ph}^\text{max}$, light stop, and
light stau scenarios are also presented, and offer very similar results relative to the $m_{\Ph}^\mathrm{mod +}$ benchmark.
Limits are determined on the product of the cross section and branching fraction $ \sigma \, \mathcal{B} ({\phi} \to \PGmp\PGmm)$
for a generic neutral boson $\phi$ at $\sqrt{s}=8$\TeV, without any assumptions on the MSSM parameters.
In this case the $\phi$ boson is assumed to be produced either in association with a $\bbbar$ quark pair or directly through gluon fusion, and sought as a single resonance with mass $m_{\phi}$. Exclusion limits are in the mass region from 115 to 500\GeV. For $m_{\phi}=500$\GeV, values $\sigma \,\mathcal{B} ({\phi} \to \PGmp\PGmm) > 4$\unit{fb} are excluded at 95\% CL for both production mechanisms. These are the most stringent results in the dimuon channel to date.

\begin{acknowledgments}
We congratulate our colleagues in the CERN accelerator departments for the excellent performance of the LHC and thank the technical and administrative staffs at CERN and at other CMS institutes for their contributions to the success of the CMS effort. In addition, we gratefully acknowledge the computing centres and personnel of the Worldwide LHC Computing Grid for delivering so effectively the computing infrastructure essential to our analyses. Finally, we acknowledge the enduring support for the construction and operation of the LHC and the CMS detector provided by the following funding agencies: BMWFW and FWF (Austria); FNRS and FWO (Belgium); CNPq, CAPES, FAPERJ, and FAPESP (Brazil); MES (Bulgaria); CERN; CAS, MoST, and NSFC (China); COLCIENCIAS (Colombia); MSES and CSF (Croatia); RPF (Cyprus); MoER, ERC IUT and ERDF (Estonia); Academy of Finland, MEC, and HIP (Finland); CEA and CNRS/IN2P3 (France); BMBF, DFG, and HGF (Germany); GSRT (Greece); OTKA and NIH (Hungary); DAE and DST (India); IPM (Iran); SFI (Ireland); INFN (Italy); NRF and WCU (Republic of Korea); LAS (Lithuania); MOE and UM (Malaysia); CINVESTAV, CONACYT, SEP, and UASLP-FAI (Mexico); MBIE (New Zealand); PAEC (Pakistan); MSHE and NSC (Poland); FCT (Portugal); JINR (Dubna); MON, RosAtom, RAS and RFBR (Russia); MESTD (Serbia); SEIDI and CPAN (Spain); Swiss Funding Agencies (Switzerland); MST (Taipei); ThEPCenter, IPST, STAR and NSTDA (Thailand); TUBITAK and TAEK (Turkey); NASU and SFFR (Ukraine); STFC (United Kingdom); DOE and NSF (USA).

Individuals have received support from the Marie-Curie programme and the European Research Council and EPLANET (European Union); the Leventis Foundation; the A. P. Sloan Foundation; the Alexander von Humboldt Foundation; the Belgian Federal Science Policy Office; the Fonds pour la Formation \`a la Recherche dans l'Industrie et dans l'Agriculture (FRIA-Belgium); the Agentschap voor Innovatie door Wetenschap en Technologie (IWT-Belgium); the Ministry of Education, Youth and Sports (MEYS) of the Czech Republic; the Council of Science and Industrial Research, India; the HOMING PLUS programme of Foundation for Polish Science, cofinanced from European Union, Regional Development Fund; the Compagnia di San Paolo (Torino); the Consorzio per la Fisica (Trieste); MIUR project 20108T4XTM (Italy); the Thalis and Aristeia programmes cofinanced by EU-ESF and the Greek NSRF; the National Priorities Research Program by Qatar National Research Fund; the Rachadapisek Sompot Fund for Postdoctoral Fellowship, Chulalongkorn University (Thailand); and the Welch Foundation.
\end{acknowledgments}

\bibliography{auto_generated}

\cleardoublepage \appendix\section{The CMS Collaboration \label{app:collab}}\begin{sloppypar}\hyphenpenalty=5000\widowpenalty=500\clubpenalty=5000\textbf{Yerevan Physics Institute,  Yerevan,  Armenia}\\*[0pt]
V.~Khachatryan, A.M.~Sirunyan, A.~Tumasyan
\vskip\cmsinstskip
\textbf{Institut f\"{u}r Hochenergiephysik der OeAW,  Wien,  Austria}\\*[0pt]
W.~Adam, E.~Asilar, T.~Bergauer, J.~Brandstetter, E.~Brondolin, M.~Dragicevic, J.~Er\"{o}, M.~Flechl, M.~Friedl, R.~Fr\"{u}hwirth\cmsAuthorMark{1}, V.M.~Ghete, C.~Hartl, N.~H\"{o}rmann, J.~Hrubec, M.~Jeitler\cmsAuthorMark{1}, V.~Kn\"{u}nz, A.~K\"{o}nig, M.~Krammer\cmsAuthorMark{1}, I.~Kr\"{a}tschmer, D.~Liko, T.~Matsushita, I.~Mikulec, D.~Rabady\cmsAuthorMark{2}, B.~Rahbaran, H.~Rohringer, J.~Schieck\cmsAuthorMark{1}, R.~Sch\"{o}fbeck, J.~Strauss, W.~Treberer-Treberspurg, W.~Waltenberger, C.-E.~Wulz\cmsAuthorMark{1}
% ...
\vskip\cmsinstskip
% ...
1:~~Also at Vienna University of Technology, Vienna, Austria\\
2:~~Also at CERN, European Organization for Nuclear Research, Geneva, Switzerland\\
% ...

\end{sloppypar}
\end{document}